\documentclass[structabstract]{aa}
\usepackage{graphicx}
\usepackage[colorlinks]{hyperref}
\usepackage{amssymb}
\usepackage{aalongtable}
\usepackage{pdflscape}
\usepackage{natbib}
\usepackage{array}
\usepackage[usenames, dvipsnames]{color}
\newcommand{\gr}{{$\gamma$-ray}}
\newcommand{\lsim}{{\lower.5ex\hbox{$\; \buildrel < \over \sim \;$}}}
\newcommand{\gsim}{{\lower.5ex\hbox{$\; \buildrel > \over \sim \;$}}}

\newcommand{\planck}{\textit{Planck}}
\newcommand{\Fermi}{\textit{Fermi}}

\begin{document}

   \title{The $\gamma$-ray emitting region in low synchrotron peak blazars}
   \subtitle{Testing self-synchrotron Compton and external Compton scenarios}

   \author{
          B. Arsioli  \inst{1,2,4} 
                  \and   
          Y-L. Chang  \inst{1,3,5}
          }
        
        \institute{
        Science Data Center della Agencia Spaziale Italiana, SSDC - ASI, Rome, Italy
        \and
        Instituto de F\'isica Gleb Wataghin, UNICAMP, Rua S\'ergio B. de Holanda 777, 13083-859 Campinas, Brazil
        \and
        Sapienza Universit\`a di Roma, Dipartimento di Fisica, Piazzale Aldo Moro 5, I-00185 Roma, Italy  
        \and
            ICRANet-Rio, CBPF, Rua Dr. Xavier Sigaud 150, 22290-180 Rio de Janeiro, Brazil
        \and ICRANet, P.zza della Repubblica 10,65122, Pescara, Italy \\
\\ 
        \email{bruno.arsioli@ssdc.asi.it, arsioli@ifi.unicamp.br}
        \email{yuling.chang@ssdc.asi.it}
        }
                  

\abstract
{}
{From the early days in $\gamma$-ray astronomy, locating the origin of GeV emission within the core of an active galactic nucleus  (AGN) persisted as an open question; the problem is to discern between near- and far-site scenarios with respect to the distance from the super massive central engine. We investigate this question under the light of a complete sample of low synchrotron peak (LSP) blazars which is fully characterized along many decades in the electromagnetic spectrum, from radio up to tens of GeV. We consider the high-energy emission  from bright radio blazars and test for synchrotron  self-Compton (SSC) and external Compton (EC) scenarios in the framework of localizing the \gr\ emission sites. Given that the inverse Compton (IC) process under the EC regime is driven by the abundance of external seed photons, these photons could be mainly ultraviolet (UV) to X-rays coming from the accretion disk region and the broad-line region (BLR), therefore close to the jet launch base; or infrared (IR) seed photons from the dust torus and molecular cloud spine-sheath, therefore far from jet launch base. We investigate both scenarios, and try to reveal the physics behind the production of \gr\ radiation in AGNs which is crucial in order to locate the production site.
} 
{Based on a complete sample of 104 radio-selected LSP blazars, with 37\,GHz flux density higher than 1\,Jy, we study broadband population properties associated with the nonthermal jet emission process, and test the capability of SSC and EC scenarios to explain the overall spectral energy distribution (SED) features. We use SEDs well characterized from radio to $\gamma$ rays, considering all currently available data. The enhanced available information from recent works allows us to refine the study of Syn to IC peak correlations, which points to a particular \gr\ emission site.
} 
{We show that SSC alone is not enough to account for the observed SEDs. Our analysis favors an EC scenario under the Thomson scattering regime, with a dominant IR external photon field. Therefore, the far-site (i.e., far from the jet launch) is probably the most reasonable scenario  to account for the population properties of bright LSP blazars in cases modeled with a pure leptonic component. We calculate the photon energy density associated with the external field at the jet comoving frame to be $\rm U'_{ext} = 1.69 \times 10^{-2}$ erg/cm$^3$, finding good agreement to other correlated works. 
} 
{}

 
 \maketitle
 
%
%

\section{Introduction}

Locating the emission site where MeV-TeV photons are produced in active galactic nuclei (AGNs) has been as an open question since the early days of \gr\ astronomy \citep{Vovk_2013_close_emission_site,Neronov2015};  one of major limitations is the angular resolution of the current generation of satellite-borne MeV-GeV and ground-based GeV-TeV observatories. Currently, we do not have  enough resolution to distinguish $\gamma$-ray substructures within the jets, even for close-by objects. The main class of AGNs detected from MeV up to tens of TeV are called blazars, and usually show extreme properties like high-power output together with short timescale  variability \citep{Aharonian_Flare_PKS_2155}, which are the main focus of studies trying to localize the \gr\ emission site. 

In summary, blazars are a particular class of jetted AGNs corresponding to the very few cases where the jet is pointing close to our line of sight \citep{PadovaniAGNwhatinname}. They are known to have a  unique spectral energy distribution (SED) often characterized by the presence of two nonthermal bumps in the log($\nu$f$_{\nu} $) versus log($\nu$) plane, extending along the whole electromagnetic window, from radio up to TeV $\gamma$ rays. Blazars are also known for their rapid and high-amplitude spectral variability. Usually, the observed radiation shows extreme properties owing to the relativistic nature of the jets, which result in amplification effects. Those objects are relatively rare. Only $\sim$ 4000 cases have been optically identified since the latest blazar surveys, 5BZcat \citet{5BZcat} and 2WHSP \citet{2WHSP}, and have been extensively studied by means of a multifrequency approach, which has cumulated  impressive dedicated databases at radio, microwave, infrared (IR), optical, ultraviolet (UV), X-ray, and $\gamma$ rays.

According to the standard picture \citep[e.g.,][]{giommisimplified}, the first peak in the log($\nu$f$_{\nu}$) versus log($\nu$) plane is associated with  the emission of synchrotron (Syn) radiation owing to relativistic electrons moving through the jet's collimated magnetic field. The second peak  is usually described as a result of inverse Compton (IC) scattering of low-energy photons to the highest energies by the same relativistic electron population that generates the Syn photons (synchrotron self-Compton model, SSC). The seed photons undergoing IC scattering can also come from outside regions (external Compton models, EC), like the accretion disk, the broad-line region (BLR), the  dust torus, and even from illuminated molecular clouds, adding extra ingredients for modeling the observed SED. 

Since the peak-power associated with the synchrotron bump tell us at which frequency ($\rm \nu_{peak}^{Syn}$) most of the AGN electromagnetic power is being released, the parameter log($\nu_{peak}^{Syn}$) has been extensively used to classify blazars. Following discussion from  \citet{padgio95} and \citet{BlazarSED}, objects with $\rm log(\nu_{peak}^{Syn}) < 14.5$, between 14.5 and 15.0, and $>$ 15.0 [Hz] are respectively called  low, intermediate, and high synchrotron peak (LSP, ISP, HSP) blazars. Some blazars whose Syn peaks reach the hard X-ray band are called extreme HSP (EHSP) blazars;  moreover, evidence for Syn peak at the MeV--GeV range are still under debate, with several cases of EHSP blazars already being studied, for example in \citet{2WHSP,Kaufmann2011,Tavecchio2011,Tanaka2014} and \citet{1BIGB-SED}. EHSP blazars are not easy to identify as they are typically faint in radio and hardly detected by current radio sky surveys; moreover, there is increasing attention from the physics community given the possibility that blazars might be associated with  astrophysical neutrinos \citep{Padovani_2016_Extreme_blazars_neutrinos,IceCube-NeutrinoTrack-2018} and with ultra-high-energy cosmic rays \citep{Resconi-2017-Blazars-UHECR-Neutrinos}. Given the broad context in which blazars play an important role for the future of astroparticle physics, studying the production site of $\gamma$ rays for the  subsample of LSP blazars may bring relevant elements for the understanding of high and very high-energy mechanisms in action for the entire blazar population. 

From current leptonic-based models, synchrotron photons and external thermal photons interacting with relativistic particles in the jet may be scattered to much higher energies, characterizing the  inverse Compton (IC) process. A simple treatment can show how this process works. In the electron frame $l^\prime$, the synchrotron photons moving along with  electrons will appear to have much lower energy, $h \nu_1^\prime \ll m_e c^2  $. In the laboratory (astrophysical source) frame $l$, the relativistic Doppler shift formula is given by
\begin{equation}
\rm h \nu_1^\prime  = \gamma  h \nu_1  \big( 1 + \beta  \cos \theta \big) \, , 
\label{lab-frame}
\end{equation}where $ \rm \theta $ is the angle between the propagation direction of photons and electrons,  $\rm \gamma=1/\sqrt[]{1-\beta^2}$ represents the Lorentz factor for the relativistic electron\footnote{This is the same as $\Gamma$, which is commonly used as a representation for the Lorentz factor associated with the bulk motion of relativistic jet-plasma.}, and $\beta = v/c$. In the electron comoving frame $l^\prime$, this angle seems much smaller $ \rm sin \theta^\prime = \frac{\sin \theta }{\gamma (1+ \beta \cos \theta ) } $ so that all photons will appear to approach in head-on collision, and Eq. \ref{lab-frame} reduces to $\rm  h \nu_1^\prime \approx \gamma  h \nu_1  \big( 1 + \beta \big)  $, since $\rm  \theta \ll 1$. Also, in $l^\prime$ frame the photon energy seems much lower ($\rm h \nu^\prime \ll m_e c^2  $) and the interaction can be treated as elastic Thomson scattering. Therefore, in $l^\prime$ frame the photon energy does not change much during the collision $\rm E_1^\prime \approx E_2^\prime$ and $\rm \nu_1^\prime \approx \nu_2^\prime$. In the AGN source frame $l$, however, photons are scattered along the direction of the relativistic electrons with $\rm  \nu_2 = \delta \nu_2^\prime \approx \delta \nu_1^\prime$, where $\rm \delta^{-1}= \gamma  \big( 1 + \beta  \cos \theta  \big)$ is the beaming factor, which reduces to $\rm \delta^{-1} \approx \gamma  \big( 1 + \beta \big) $ so that we have $ \rm \nu_2 = \nu_1 \big[ \gamma \big( 1 + \beta \big) ]^2 $. In the relativistic limit, $ \rm \beta \approx 1$, and the frequency associated with the upscattered photon  follows as $\rm \nu_2 \approx 4 \ \gamma^2 \ \nu_1 $.


Therefore, an important conclusion is that photons scattered by relativistic electrons gain energy with a $\rm \gamma^2$ factor:  $\rm E_2 \propto \gamma^2 E_1$. Naturally, the luminosity $\rm L_{IC}$ of the inverse Compton component depends on the photon density $n_{ph}$ available for up-scattering via the IC process. In the SSC model only photons generated by the synchrotron process itself may build up the available $n_{ph}$. In addition, the external contribution from thermal emission regions can be significant sources of low-energy photons, characterizing the  EC models. In both cases we have $\rm L_{IC} \propto n_{ph} \gamma^2 E_{1} $ \citep{RadiativeProcessAstrophysics}.    

As is known, the synchrotron emission can extend up to hard X-rays, and in some extreme cases can even peak in this region. When the photon energy reaches a  level that is similar  to the electron mass, the condition $h \nu^\prime \ll mc^2$ is not valid in the electron's frame, and Klein--Nishina effect (described by applying quantum electrodynamics to the scattering process) acts to reduce the electron-photon cross section with respect to the case of classical Thomson scattering ($\sigma_T$). Therefore, the IC scattering might becomes less and less efficient for seed photons with the highest energies (e.g., $\rm \sigma_{KN}  /  \sigma_T  \approx 0.5 $ at E\,=\,300\,KeV), which influence the spectral energy distribution of blazars at very high energies E\,$>$\,100\,GeV and manifest as a strong break (steepening) in \gr\ emitted power. In fact, if the electron energy distribution follows $\rm N_{(E)}=k E^{-p}$, the scattered IC spectrum will also be a power law with spectral index $\rm \alpha = (1-p ) / 2$.

Although a pure leptonic IC process is well established as the mechanism that produce the second bump observed on the blazar's SED, there is still open debate on alternative scenarios like the ones considering hadronic plus leptonic components \citep{Bottcher2013,Cerruti2011,Cerruti2017}. In addition, the location and AGN environment dependences associated with the production of  $\gamma$ rays  are still unclear.  Probing such information demands a set of multifrequency measurements together with model-dependent tests, as we discuss  below. Given the many identified $\gamma$-ray sources, there is still a great deal of  room to explore issues like variability (comparing the behavior at low and high energies) and probing the far end of SED at E\,$>$\,10\,TeV with the upcoming generation of Cherenkov telescope arrays \citep[CTA, ][]{CTA50h}.

In this work we focus on modeling low synchrotron peak (LSP) blazars making use of a complete sample of radio-loud blazar AGNs described in details by \citet{PlanckEarlySEDs}. It consists of 104 northern and equatorial sources with declination greater then -10$^o$, flux density at 37\,GHz exceeding 1\,Jy as measured with the Mets\"ahovi radio telescope. All 104 sources have been detected between 30\,GHz and 857\,GHz by the \planck\ mission \citep[Planck Catalogue of Compact Sources PCCS, ][]{PCCS-Planck-Catalog} most of which were previously known. With the addition of PCCS data, many radio-bright blazars gained a better multifrequency description for their synchrotron (Syn) component, and here are referred to  as radio-Planck sources. 

It is important to note that the vast majority of these sources (103) are legitimate LSP blazars (two cases at the border line, $\nu_{peak} \approx$10$^{14.5}$\,Hz, BZQJ\,0010+1058 and BZBJ\,0050-0929); only one bright HSP blazar (BZBJ\,1653+3945) was removed or properly highlighted during the preparation of following studies. Out of those 104 sources, 83 have a confirmed \gr\ counterpart in at least one of the \Fermi-LAT \citep{FermiLAT} catalogs 1FGL, 2FGL, and 3FGL \citep[][]{1FGL,2FGL,3FGL} and another 16 had their $\gamma$-ray spectrum recently described by \cite{paper1}, who search for new $\gamma$-ray emitting blazars following the same approach as \cite{1BIGB}. We note that their study was based on a dedicated \textit{Fermi}-LAT analysis showing that many of the previously \gr\ undetected LSPs are actually detectable when integrating over 7.5 years of observations.

The online SED builder tool\footnote{The SED builder is an online tool dedicated to multifrequency data visualization, together with fitting routines useful for extracting refined scientific products. Provided by the Space Science Data Center (SSDC): \url{http://www.ssdc.asi.it}} was used in previous work to compile and fit all available multifrequency data \citep{paper1} that we now use for current analysis. This included relevant microwave flux measurements coming from the Planck mission, the new $\gamma$-ray data-points from the \Fermi-LAT dedicated analysis, and extra UV to X-ray observations from Swift. From there, fitting parameters were extracted to describe the observed peak-frequency log($\nu_{peak}$) and peak-brightness log($\nu$f$_{\nu}$) for both Syn and IC bumps. We now use those measurements to gain further insight on the population properties of LSP blazars, calculating parameters like the Lorentz factor associated with relativistic electrons in the jet, the product B$\delta$ ($\delta$ stands for the beaming factor), the luminosity associated with Syn and IC peaks, and the external photon field energy density (U$_{ext}$) calculated when assuming an EC models.

\section{LSPs jets and nonthermal emission mechanism}

As argued in the literature \citep{MeVPeaked2017,why-no-detection} LSP blazars with $\rm \nu_{peak} < 10^{13.4}$Hz may show a typical inverse Compton peak below 0.1\,GeV, and thus out of the \Fermi-LAT sensitivity bandwidth at 0.1--500\,GeV. In the case of LSP blazars, we might probe only  the very end of the IC component, and this is why  a considerable percentage of LSPs ($\sim$20\%) had no counterpart in the latest \Fermi-LAT catalogs (1FGL, 2FGL, and 3FGL). The relation between Syn and IC peaks is explored in \citet{BlazarSED,Gao2011_LsyLIC,Zhang2012}, with \citet{Senturk2013} showing a correlation between peak frequencies, since $\nu^{IC}_{peak}$ is decreasing with respect to HBL-LBL-FSRQ. There is a clear connection between the distributions of $\rm log(\nu^{Syn}_{peak})$ and $\rm log(\nu^{IC}_{peak})$ when a complete sample of LSPs is considered, such that a characteristic peak ratio ($\rm PR=log(\nu^{IC}_{peak}/\nu^{Syn}_{peak})$ is very suitable for describing the average relation between their distributions \citep[$\rm PR\approx 8.6$,][]{paper1}. However, when taken case by case, they show that an intrinsic and direct relation between peak frequencies is nontrivial and most probably highly dependent on its SSC or EC dominance nature and variability.

Intrinsic jet properties like the beaming factor ($\delta$) and the dominant IC regime (either synchrotron self-Compton,  SSC, or external Compton, EC) may in fact have a strong influence on the $\gamma$-ray variability, and also affect the \Fermi-LAT detectability of a few radio-loud blazars. \cite{RadioLoudDoppler-Beamed} have shown that the \gr\ sources detected during the first three months of \Fermi-LAT operations are on average the ones associated with the highest apparent jet speeds (based on radio measurements with the Very Large Baseline Array, VLBA) and therefore the most powerful accelerators with the highest $\rm \delta$ values. 

In a simple SSC scenario \citep{Maraschi1992_SSC1,Marscher1996_SSC2} the intensity boosting factor scales as $\delta ^{ 3+\alpha}$, where $\alpha$ is the spectral index given that the flux scales as S$_{\nu} \propto \nu^{-\alpha}$. When considering typical blazar SEDs in the S$_{\nu}$ versus $\nu$ plane, the spectrum tends to be flat at radio frequencies and steep in $\gamma$ rays.  As a consequence, the intensity boosting is more pronounced in $\gamma$ rays than in the radio bands, and thus the \gr\ detection of faint sources is favored during flaring episodes. 

Considering the mechanism involved for the IC scattering, additional photon fields could be present and even dominant with respect to synchrotron photons \citep{Dermer,EC-model1}. For instance, the jet might interact with external photons produced by the accretion disk, reflections, and IR thermal emission from surrounding gas clouds and dust torus. In such scenarios, a considerable amount of the \gr\ emission would be produced by Compton scattering of those external photons which is associated with an additional boosting factor $\delta^{1+\alpha}$ enhancing the IC intensity.

Given that nearly all radio-Planck sources are now well described in $\gamma$ rays, we study their population properties to probe the leading emission mechanism, either SSC or EC, looking for hints to locate the \gr\ emitting region in LSP blazars. 

\section{Comments on synchrotron self-Compton  and external Compton  scenarios}

Identifying the dominant IC mechanism can help to locate the $\gamma$-ray emitting region and to better understand its underlying physics. If the IC emission is dominated by EC process and happens close to the black hole (0.1--1 pc distance, embedded in the BLR)  it could explain the observed $\gamma$-ray short timescale variability of a few hours \citep{Aharonian_Flare_PKS_2155,Pittori2018}. In this case, since optical-UV BLR photons would be available for up-scattering, there should be a strong correlation between $\gamma$-ray and optical-UV flares. However, an alternative scenario considers that the $\gamma$-ray emission originates  farther from the black hole (BH) at distance $\gg$1 pc. In this case, an IR photon field generated by the molecular clouds and dust torus (through reprocessing radiation from the accretion disk, or even by illumination from the jet synchrotron emission itself) are possible dominant sources of seed photons for up-scattering to higher energies \citep{Breiding-2018-Sheath}. 

Both SSC and EC scenarios with $\gamma$ rays originating far from the BH (out of the BLR region) demands that the jet structure should be a very narrow opening ($\approx$0.8 pc scale) or have N substructures ($\approx$0.8/$\sqrt N$ pc), as invoked by multicomponent scenarios, to reconcile with the short timescale variability observed in the GeV--TeV band \citep{Agudo2011}. 

There are indeed plenty of arguments supporting the far-site emission. One is related to radio-mm observations with the Very Long Baseline Array (VLBA) which shows radio-mm variability \citep[at a distance on the order of $~$12--14 pc from the BH, for the BL Lac AO 0235+164 and OJ 287,][]{Agudo2011,Agudo2013} to be correlated in time with $\gamma$-ray flares, and therefore supposed to happen in the same site. If this is true, there remains the question of how  all the power gets transferred so efficiently farther away from the BH to produce the VHE component we observe. In addition, BL Lacs are usually dominated by nonthermal emission along the whole spectrum, with no trace of disk or dust torus thermal component. 

\cite{Agudo2013} has interpreted the far emission site for BL Lacs AO 0235+164 and OJ 287 in the framework of SSC scenario since no other evident photon field is present for up-scattering other then the nonthermal synchrotron photons. However, similar work from \citet{Ackermann2012_Flaring} for the same object and flaring event (AO 0235+164, 2008) describes the IC component within an EC scenario, assuming the external photon field is dominate by IR photons from the surrounding dust torus, which are up-scattered to HE. An extra IR component could also be present and dominant as a consequence of illumination and sublimation of the molecular cloud (MC) torus in a spine-sheath geometry as described by \citet{Breiding-2018-Sheath} and \citet{MacDonald-2015-Sheath}. We should not avoid  mentioning several works, such as \citet{Close-Emission-Site-2014} and \citet{Neronov2015} in favor of a close-site emission at the vicinity of the AGN supermassive black hole.

Even though we do not have evidence of MC or dust torus thermal components in BL Lacs, we should note that the nonthermal jet synchrotron emission is beamed and dominant because our observer frame is watching a relativistic jet pointing close to our line of sight. Thermal components from the MC and dust torus could well be present, nevertheless swamped by the jet beamed emission. Given the relativistic nature of the jet, even a relatively faint photon field with energy density $\rm U_{ext}$ would be boosted in the jet's ``$\prime$'' reference frame $\rm U^\prime_{ext} \propto U_{ext} \Gamma^2$ \citep{Ghisellini_Madau_1996}. This could become much more relevant or even dominant with respect to the Syn photon field produced by the jet itself, and strongly depending on the Lorentz factor ($\Gamma$).  

The discussion surrounding the $\gamma$-ray emitting region in AGNs is ample, and interpretations are always subject to multiple free parameters that can be fine-tuned for different scenarios. We try to contribute to that understanding by studying general properties of the radio-Planck sample as a fair representation of powerful LSP blazars.

\section{Lorentz factor of  jet's  relativistic electrons}

Assuming a homogeneous SSC to describe the blazar SED, high-energy photons are generated by the up-scattering of low-energy Syn photons due to their interaction with relativistic electrons from the jet. In this scenario, a single population of relativistic electrons is then responsible for the entire SED, resulting in a strong correlation between the Lorentz factor of the electrons $\rm \gamma_{peak}$ emitting at the peak of the Syn component, and the peak frequency from the Syn and IC components:

\begin{figure}[]
   \centering
    \includegraphics[width=1.0\linewidth]{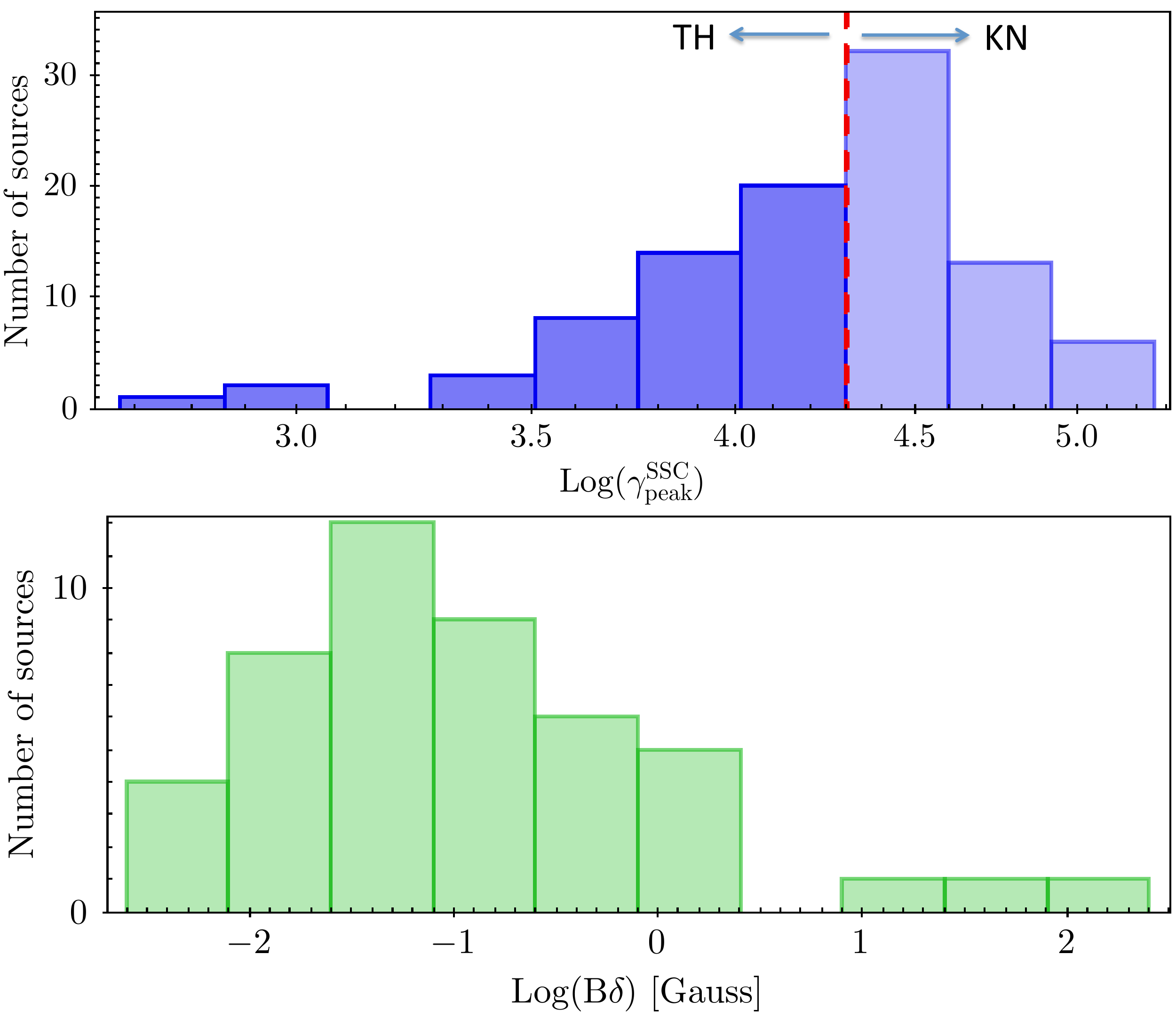}
     \caption{Top: Distribution of Lorentz factor $\rm \gamma^{SSC}_{peak}$ for electrons emitting in the peak of the Syn component (eq. \ref{gamma-SSC}). The red dashed line shows the limit for $\rm \gamma^{SSC}_{peak}$ values where the transition from TH to KN scattering regime occurs. Bottom: Distribution of the $\rm B \delta$ parameter for LSP sources assuming a SSC model, and considering only sources with $\rm \gamma^{SSC}_{peak}< 2 \times 10 ^4$ (under TH regime).}
      \label{LorentzDist}
\end{figure}

\begin{equation}
\gamma_{peak} \simeq \sqrt{3/4 \times \nu^{IC}_{peak}/ \nu^{Syn}_{peak}}  \  
\label{gamma-SSC}
\end{equation}
However, this trend is valid only under Thomson regime (TH) of the IC scattering, that is for $\rm \gamma_{peak} < 2 \times 10^4$, where the transition to the Klein--Nishina (KN) regime occurs. We use eq. \ref{gamma-SSC} to calculate $\rm \gamma^{SSC}_{peak}$ for all 99 sources with Syn and IC parameters available (Table \ref{tableRadioPlanck}), and plot its distribution in fig. \ref{LorentzDist}, top. The histogram shows a Gaussian distribution with slightly negative skewness, and characterized by mean $\rm \langle log(\gamma^{SSC}_{peak}) \rangle =4.24 \pm 0.05$. Therefore, half of the sample have $\rm \gamma^{SSC}_{peak} > 2 \times 10^{4}$, which is in tension with the fact that we are dealing with bright LSPs. 
Apart from a single HSP (Mrk501), all sources have $\rm \nu^{Syn}_{peak}< 10^{14.6}$ Hz and the transition from TH to KN regime is only expected for $\rm \nu^{Syn}_{peak}> 10^{14.7}$\,Hz in the case of a single-zone SSC model.

In a simple single-zone self-synchrotron model, the $\rm \nu^{Syn}_{peak}$ can be written in terms of jet magnetic field (B), beaming factor ($\delta$), and peak Lorentz factor ($\rm \gamma_{peak}$). As discussed in \cite{BlazarSED}, assuming an emitting region of size R$\sim$10$^{15}$cm\footnote{Given that $\rm R < \frac{c \Delta t \delta}{(1+z)}$, assuming z on the order of 1.0, and $\rm \delta \sim 10-20$, with characteristic variability timescale of a few days ($\sim$5$\times$10$^5$ s), jet length is on the order of R$\,\approx\,$10$^{15}$ cm.}, and a log-parabola to describe the distribution of Lorentz factor: $ \rm n_{(\gamma)}\,=\,k\times10^{r \ log(\gamma/ \gamma_{peak})^2}$ (curvature parameter r\,=\,2.0, $\gamma$ ranging from 10$^2$ to 6$\times$10$^5$, and electron density of $\rm \sim 1.0 \, cm^{-3}$; \citealt{Tramacere2010_SSC_ECplot}), we have 
\begin{equation}
\nu^{Syn}_{peak} = 3.2 \times 10^6 ( \gamma_{peak})^2 B \delta /(z+1)  \ , 
\label{gamma-syn}
\end{equation}
which is valid up to $\rm \nu^{Syn}_{peak} \approx 10^{14.7}$ Hz, where the transition to KN scattering regime occurs. Following the discussion from \cite{BlazarSED}, $\rm \gamma^{SSC}_{peak} \approx \gamma_{peak}$ under TH regime, therefore we use eq. \ref{gamma-syn} to calculate the B$\delta$ parameter as B$\rm \delta=\nu^{Syn}_{peak}(1+z)/(3.2\times10^6 \times (\gamma^{SSC}_{peak})^2$), only for the subsample of 48 sources having $\rm \gamma^{SSC}_{peak}< 2 \times 10 ^4$. These 48 sources are the ones under TH regime if we assume a single-zone SSC model. We plot the B$\delta$ distribution in Fig. \ref{LorentzDist} (bottom) which peaks at $\langle B\delta\rangle$=0.066$^{+0.020}_{-0.015}$ gauss. This is also in tension with the expected value for the B$\delta$ parameter for blazars, which is usually assumed to be $\rm \langle B\delta\rangle_{expected}=10$ gauss, with beaming factor $\delta$ on the order of $\sim$20 \citep[ranging from 5 to 35 for LSP blazars,][]{Kang2014_Lorentzfactor} and B on the order of $\sim$0.5 gauss \citep[ranging from 0.3 to 1.5 gauss,][]{Tramacere2010_SSC_ECplot}. Most probably, the $\gamma^{SSC}_{peak}$ values that we have calculated from eq. \ref{gamma-SSC} are highly overestimated, leading to low B$\delta$ values. Therefore a simple single-zone SSC model seems insufficient to account for the overall SEDs observed  for LSP blazars.


\subsection{Tramacere plane: log($\gamma^{SSC}_{peak}$) versus log($\nu^{Syn}_{peak}$) }

\begin{figure}[h]
   \centering
    \includegraphics[width=1.0\linewidth]{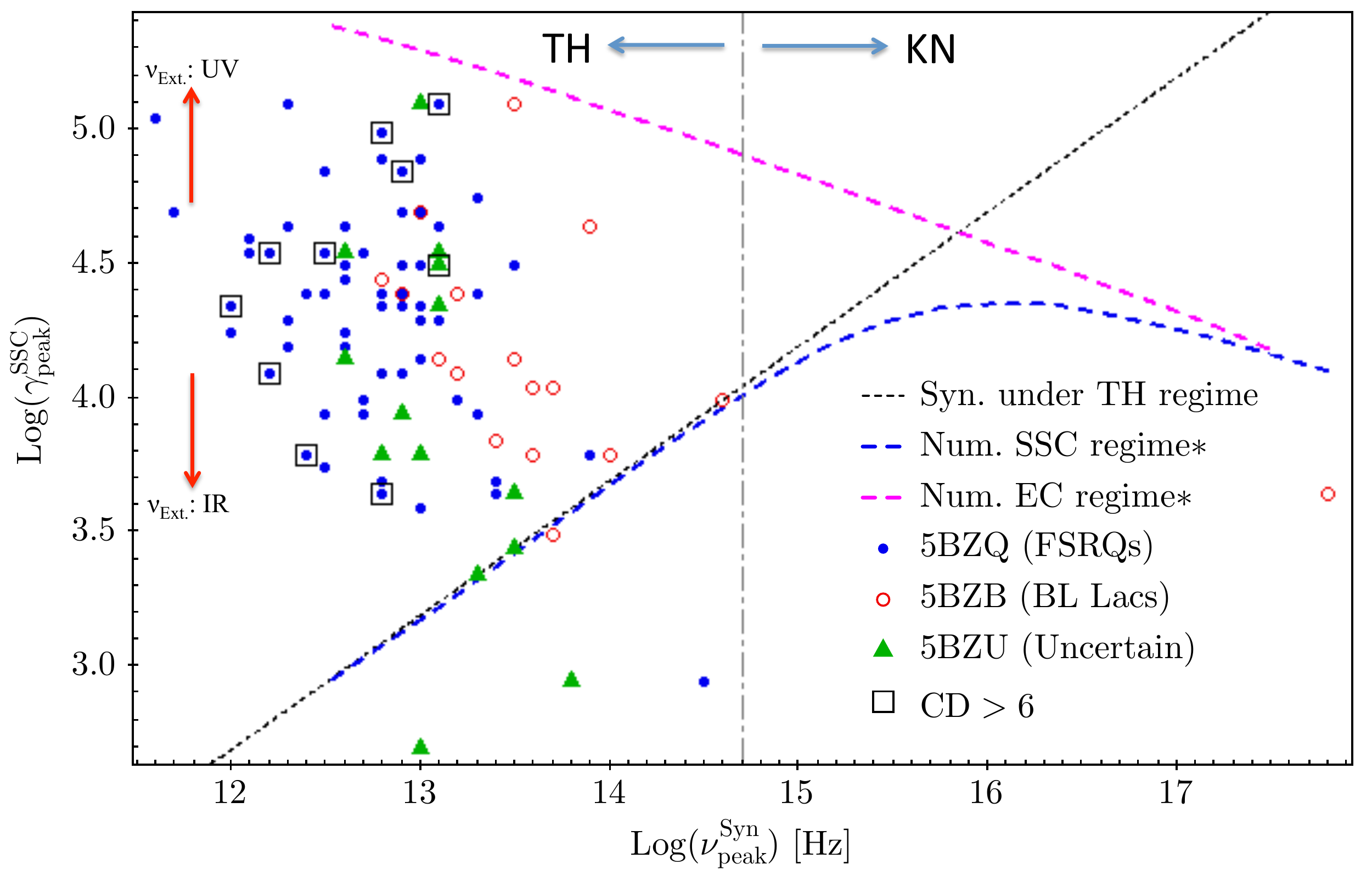}
     \caption{ The $\rm log(\gamma^{SSC}_{peak})$ vs. $\rm log(\nu^{Syn}_{peak})$ plane. Sources are divided according to their classification in the 5BZcat: blue  circles for 5BZQs (FSRQ), red empty circles for 5BZBs (BL Lacs), and green triangles for 5BZUs (unclassified). Blue and purple dashed lines correspond to $\gamma^{SSC}_{peak}$ calculated from numerical simulations considering SSC and EC scenarios incorporating the transition from TH and KN regimes. The vertical dot-dashed line indicates the $\nu^{Syn}$ domains for which TH and KN regimes apply. The black dashed line represents eq. \ref{gamma-syn} using B$\delta$/(1+z)$\approx$1.3 gauss, with no correction for the TH to KN transition.}
      \label{nuICgamma}
\end{figure}

\cite{Tramacere2010_SSC_ECplot} proposes the use of the log($\gamma^{SSC}_{peak}$) versus log($\nu^{Syn}_{peak}$) plane to better understand the dominant emission mechanism in blazars (either SSC or EC) for individual sources and populations, as also mention by \cite{BlazarSED}. In fig. \ref{nuICgamma} we show the radio-Planck  sources in this plane, with $\rm \gamma^{SSC}_{peak}$ values estimated using eq. \ref{gamma-SSC}, directly from the Syn and IC peak-power parameters measured from fitting the SEDs case by case (Table \ref{tableRadioPlanck}). 

In this plane, the blue dashed line (extracted from \citealt{Tramacere2010_SSC_ECplot}, their Fig. 3) represents a SSC numerical model which incorporates the TH to KN transition, and therefore corrects for the decreasing e$^-$+$\gamma$ cross-section  which reduces the efficiency of IC scattering and affects the cooling time of relativistic electrons in the jet. The black dashed line represents the synchrotron emission, simply plotting eq. \ref{gamma-syn} for the SSC model with no correction on the TH to KN transition (assuming R$\sim$10$^{15}$cm and B$\delta$/(1+z)=1.3 to match with the SSC-TH estimate from the numerical modeling). The purple dashed line (\cite{Tramacere2010_SSC_ECplot}, also from their Fig. 3) represents a numerical model for the EC regime assuming benchmark values for the jet parameters: R$\sim$10$^{15}$cm, a log-parabola to describe the distribution of Lorentz factor ($\gamma$) of the jet's relativistic electron, 
assuming a dominant UV external photon field produced by the accretion disk (modeled as a blackbody with T profile having innermost T of $\approx$10$^5$ K), and assuming an extra component reflected by the BLR toward the jet, with efficiency $\tau$=10\%. Those models are described and applied in a series of works: \citet{tramacere2003,MassaroTramacere2006_codeIII,TramacereThesis2007,tramacere2009}.

We separate sources according to their classification in the 5BZcat catalog (BZBs, BZQs, Uncertain types, \cite{5BZcat}), and also mark cases with the highest Compton Dominance (CD) values (CD\,$>$\,6.0 to select the top 10\% of sources). Most sources cluster in the region above the blue dashed line, meaning they are mainly out of the SSC domain. This region is characteristic of blazars where there might be an external photon field ranging from IR to UV playing an important role. 

In conclusion, an EC mechanism under the TH scattering regime should be more suitable to study those sources. The Tramacere plane then gave us an overview on the dominant IC mechanism in play for bright LSP blazars, and also shows that there is no significant differences (data clustering) with respect to the $\gamma^{SSC}_{peak}$ parameter depending on blazar type or Compton dominance.


\subsection{Assuming a dominant EC scenario}
\label{ECsection}

If we assume that a source can be described via the EC model under the TH regime, the frequency associated with the IC peak ($\rm \nu^{EC}_{peak}$) should be well described by \citep{BlazarSED} 
\begin{equation}
\rm \frac{ \nu^{EC}_{peak}}{\nu^{ext}_{peak} \Gamma}= \frac{4}{3} \big(\gamma_{peak} \big)^2  \frac{\delta}{(1+z)}
\label{ec-eq}
,\end{equation}
where $\gamma_{peak}$ is the Lorentz factor associated with jet electrons emitting in the peak of the synchrotron component (see eq. \ref{gamma-SSC}) and $\rm \nu^{ext}_{peak} $ is the peak frequency associated with the external photon field in the rest frame from the emitting zone (either accretion disk, BLR, MC, or dust torus). When $\rm \nu^{ext}_{peak} $ is multiplied by the bulk Lorentz factor $\Gamma$ associated with the relativistic outflow, it transforms this frequency to the jet rest frame. We  use the notation $\rm \gamma^{EC}_{peak}$ to represent $\rm \gamma_{peak}$ when assuming an EC scenario. 

\begin{figure}[h]
   \centering
    \includegraphics[width=1.0\linewidth]{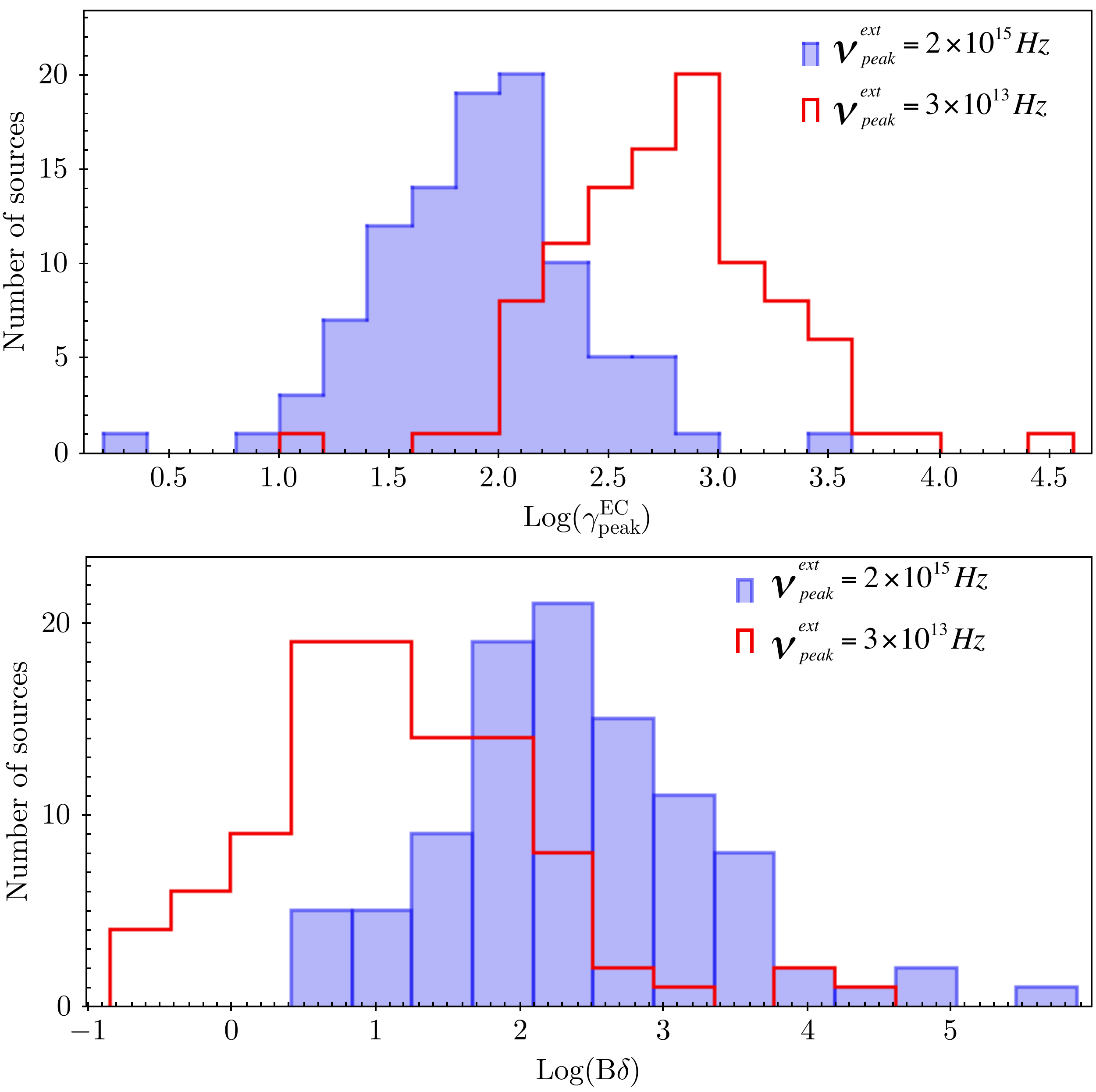}
     \caption{Top: Distribution of $\rm log(\gamma^{EC}_{peak})$ for the radio-Planck sample based on eq. \ref{ec-eq} when assuming $\rm \Gamma=20$, and for two different external photon fields: dust torus photons peaking at IR $\rm \nu^{ext}_{IR} =3 \times 10^{13} Hz $ (in red);  near-UV photons from the BLR region, peaking close to $\rm Ly_{\alpha}$, $\rm \nu^{ext}_{UV} =2 \times 10^{15} Hz $ (full indigo bars). Bottom: Corresponding distribution of $\rm log(B\delta$) calculated from eq. \ref{gamma-syn} for $\rm \gamma_{peak} = \gamma^{EC}_{peak}$, and considering   $\nu^{Syn}_{peak}$ and $\nu^{IC}_{peak}$ from Table \ref{tableRadioPlanck}.}
      \label{gamma-EC}
\end{figure}

We use eq. \ref{gamma-EC} to calculate $\rm \gamma^{EC}_{peak}$ for all sources with available IC data, considering $\rm \nu^{IC}_{peak}$ ($\rm \equiv \nu^{EC}_{peak} $) as reported in Table \ref{tableRadioPlanck}. To perform this calculation we assume the Doppler (beaming) factor $\rm \delta=[\Gamma(1-\beta cos \theta)]^{-1} \approx \Gamma$ \citep{Dermer2015_blazarParadigm} valid for sources observed close to the line of sight, $ \rm \theta < 5^{\circ}$. 

We assume $\langle \delta \rangle$$\approx$\,20$\pm$2, following \cite{Kang2014_Lorentzfactor}, which presents a list of $\delta$ parameter for 15 bright LSPs, as estimated from the model constrained by SED fitting\footnote{The adopted model considers an EC leptonic scenario, assuming external photon fields from BLR (UV), molecular, and dust torus (IR), with the last resulting in better fittings.} and in agreement with estimates from radio variability and brightness temperature (confirming early measurements made by \cite{Jorstad2005}). Also, \cite{Saikia2016} introduced a new independent method based on the optical fundamental plane of black hole activity\footnote{The method \citep{Saikia2015} is based on the fundamental plane of black hole activity in X-rays. The proposed ``optical fundamental plane of BH activity'' relies on the OIII forbidden-line intensity (independent of beaming and viewing angle) as a tracer for the accretion rate instead of the X-ray flux, which is heavily contaminated by a nonthermal jet component in blazars.} to estimate the $\Gamma$ distribution, showing a valid range from 1 to 40, with N$_{(\Gamma)} \propto \Gamma^{-2.1 \pm 0.4} $, or an even more restrictive range with $\Gamma$  between 15 and 30 \citep{Close-Emission-Site-2014}, as deduced from a study of $\gamma$-ray flares, with a multifrequency approach and testing EC scenarios.

There are two different setups that are important to consider, and that are related to the photon-frequency ($\nu$) associated with the external photon field. The first one assumes that seed photons originate mainly from the dust torus. This view is supported by \cite{Cleary2007_SpitzerTorus} who deduced from observations with {\it Spitzer} that the torus may heat up to 150--200 K by absorbing accretion disk radiation and emitting like a blackbody, and therefore with dominant IR emission peaking at $\rm \nu^{ext}_{IR} =3 \times 10^{13} Hz $. There is a similar scenario where the illumination and sublimation of molecular clouds, owing to synchrotron jet emission in a spine-sheath geometry \citep[][]{Breiding-2018-Sheath}, could also play important role in producing a dominant IR photon field. In the second setup  the external photon field originates from the BLR and accretion disk regions, with dominant emission peaking close to Ly$_{\alpha}$ in near UV, at $\rm \nu^{ext}_{UV} =2.0 \times 10^{15} Hz $ \citep{TravecchioGhisellini2008,Ghisellini2008_EC-BLR}. 

In Fig. \ref{gamma-EC} (top) we plot the distribution of $\rm log(\gamma^{EC}_{p})$ for the radio-Planck sample, which leads us to the following conclusion. When using an EC model with external photon field ranging from UV to IR, and assuming $\langle \delta \rangle \approx \langle \Gamma \rangle$\,=\,20, almost all sources are under the TH scattering regime. This is in agreement with expectations since the radio-Planck sample is dominated by LSP blazars, $\rm \nu^{Syn}_{peak} <10^{14}$\,Hz. We find $\rm \langle log(\gamma^{EC}_{peak}) \rangle$ ranging from 2.80$\pm$0.05 to 1.93$\pm$0.05 depending on the external photon field, UV and IR, respectively. Compared to $\gamma^{SSC}_{peak}$ values calculated when assuming a simple SSC model (Fig. \ref{LorentzDist}), we see that  $\gamma^{SSC}_{peak}$ is highly overestimated by almost two orders of magnitude in any scenario.

As mention previously, eq. \ref{gamma-syn} is only valid under the TH regime. Therefore, we recalculate the B$\delta$ parameter according to B$\rm \delta=\nu^{Syn}_{peak}(1+z)/(3.2\times10^6 \times (\gamma^{EC}_{peak})^2$), which now applies to the subsample of 98 sources having $\rm \gamma^{EC}_{peak}< 2 \times 10 ^4$. We plot the B$\delta$ distribution in Fig. \ref{gamma-EC} (bottom), which peaks at $\rm \langle log(B\delta)\rangle = 2.72 \pm 0.09$ for the UV external field and at $\rm \langle log(B\delta) \rangle = 0.99 \pm 0.09$ for the IR external field. Therefore, assuming $\rm \langle \delta \rangle = 20$ we get an estimate for the magnetic field in the jet $\rm \langle B_{UV} \rangle = 26.2$ gauss, and  $\rm \langle B_{IR} \rangle = 0.48$ gauss.  In particular, the estimate for $\rm \langle B_{UV} \rangle $ is not consistent with the constraints from SED fitting when assuming an emission site within the BLR \citep{Gang2013_LocationGamma}, owing to underestimated $\rm \gamma^{EC}_{peak}$ values. However,  the estimate for $\rm \langle B_{IR} \rangle$ is in good agreement with expectations from SED fitting from \cite{Kang2014_Lorentzfactor} for $\gamma$-ray emission out of the BLR region (far-site) at a distance $\gg$ 0.1 pc  from the BH. 


This suggests that an IR external photon field might be the dominant driver in the EC scenario for the population of bright LSP blazars, also in agreement with findings from \cite{Fermi2015_PKS_lensed}. One important aspect to note is that the energy density (U) from external photon fields are boosted in the jet's comoving frame ``$\prime$'' according to $\rm U^\prime_{ext} = U_{ext} \Gamma^2$ \citep{Sikora2009}, therefore strongly dependent on the jet's bulk Lorentz factor $\Gamma$ and accounting for $\rm U^\prime_{ext}$ being dominant with respect to the self-synchrotron photon field. 

If we assume an UV external photon field from the BLR region, forcing the magnetic field to $\rm 0.5 < B < 2.0$ gauss as expected from SED fitting derived from \cite{Gang2013_LocationGamma}, it may lead to highly overestimated $\gamma^{EC}_{peak}$ values, as also reported by \cite{BlazarSED}. 
In fact, when relaxing the value associated with $\langle \delta \rangle$, it is possible to adjust UV dominant scenarios for some individual sources, and that is a known degeneracy associated with the B$\delta$ parameter.

 
\subsection{Syn versus IC luminosity correlation}

\begin{figure}[]
   \centering
    \includegraphics[width=1.0\linewidth]{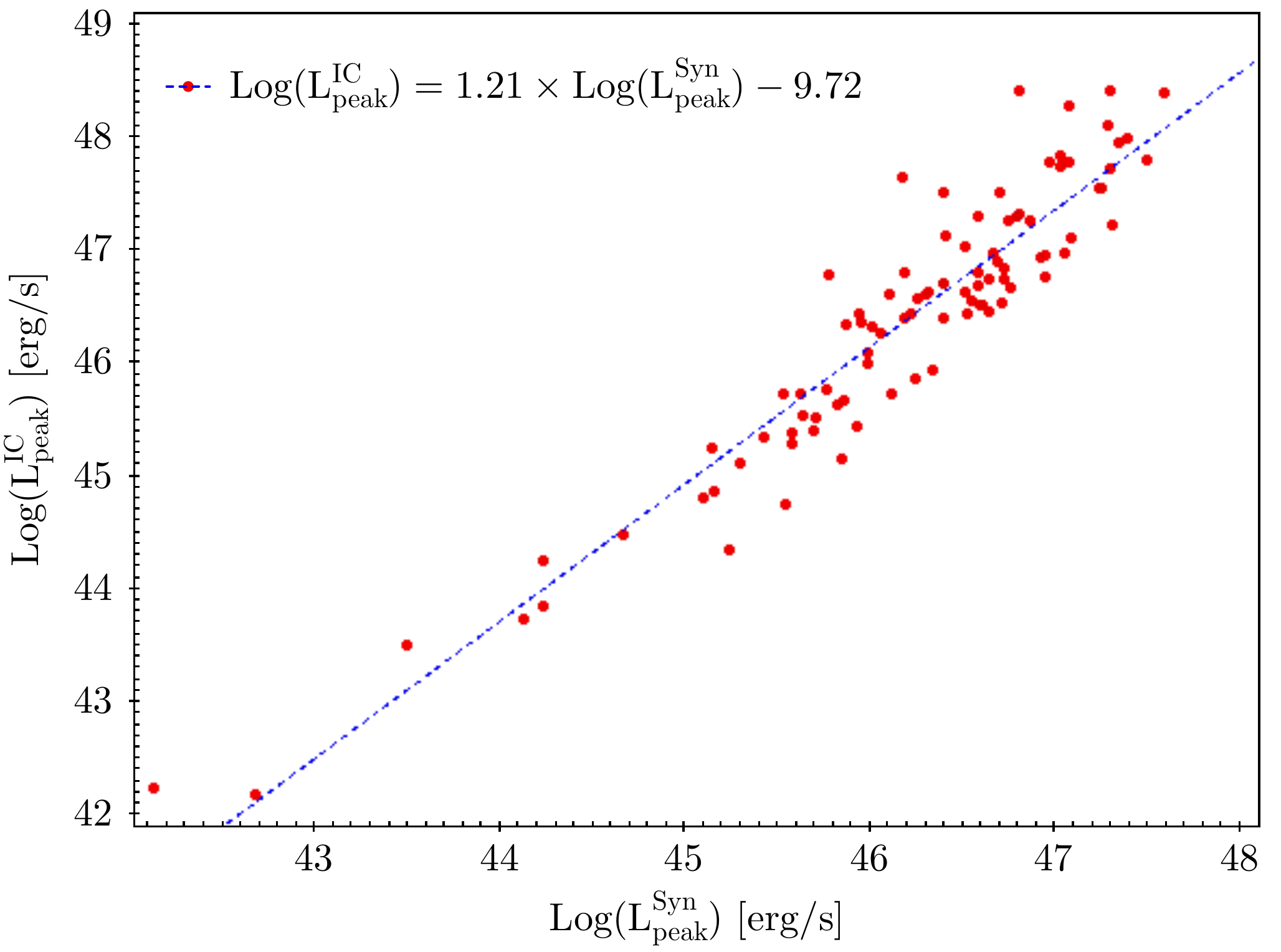}
     \caption{Syn vs. IC luminosity plane showing a tight correlation that extends for seven decades in luminosity. Blue dashed line is a linear fit to the data.}
      \label{luminosity}
\end{figure}

We have calculated the Syn and IC peak luminosities based on the flux density $\nu$f$_{\nu}$ [ergs/cm$^2$/s] measurements listed in Table \ref{tableRadioPlanck}. Luminosity is given by

\begin{equation}
\rm L = \frac{4 \pi d^2_L  \ \nu f_{\nu}}{(1+z)^{1-\alpha}} \equiv  4 \pi d^2_L  \ \nu f^{peak}_{\nu}
,\end{equation}
where $\rm d_L$ is the luminosity distance calculated based on $\Lambda$CDM cosmology (with H$_0$=67.3 Km/s/Mpc, $\omega_{\Lambda}$=0.685, $\omega_K$=0, and $\omega_M$=0.315, \cite{LCDMpar}). Given that we calculate the luminosity at the Syn and IC peaks measured from the SEDs in the $\nu$ versus $\nu$f$_{\nu}$ plane, the photon spectral index is $\rm \Gamma^{IC,Syn}_{peak}=2.0$; therefore, $\rm \alpha=(\Gamma-1)=1.0$, and the K-correction term simplifies to $\rm (1+z)^{1-\alpha}=1.0$. As seen from Fig. \ref{luminosity} the scatter in the $\rm log(L_{Syn})$ versus $\rm log(L_{IC})$ plane is very tight, holding along seven decades in luminosity, with a strong Pearson correlation coefficient of 0.94. The correlation is described by

\begin{equation}
\rm log(L^{IC}_{peak}) = 1.21 \times log(L^{Syn}_{peak}) -9.72
\end{equation}

This relation was also probed by \cite{Gao2011_LsyLIC} using an early data release from \Fermi-LAT after three months of observations, plotting the total Syn against IC luminosities. Their relation between $\rm log(L_{IC})$ and $\rm log(L_{Syn})$ had a slope of 1.1, similar to our value. Although their correlation coefficient is much lower, 0.58 (owing to  larger uncertainties in the $\gamma$-ray band, especially because of low \Fermi-LAT exposure and its early detector calibration at the time), the agreement is remarkably good.      

In the luminosity plane (Fig. \ref{luminosity}) we are most likely probing the mean behavior of both Syn and IC emission. Especially for the $\gamma$-ray band, the spectral data points were calculated integrating over a few years of \Fermi-LAT observations; therefore, short flaring states (day-week scale) are smoothed and the IC luminosity we plot is a fine representation of the mean emitted power.

The correlation we see at the luminosity plane is probably related to a constant ratio between external photon field ($\rm U'_{Ext}$) and the magnetic field ($\rm U'_{B}$) energy densities in the jet comoving frame. Assuming an EC scenario in this case, this correlation could be taken as observational evidence of the established balance between a dynamic radiative-drag and the magnetic energy density. On the one hand, the radiative-drag is induced by the jet interaction with a boosted external photon field $U'_{Ext}$, as discussed in \citet{Moderski-Radiation-Drag-2003} and \citet{Madejski-Radiation-Drag-1999}, which is directly connected to the loss energy mechanism for the relativistic electrons (cooling) even imposing limitations to the jet's Lorentz factor ($\Gamma$). On the other hand, following \citet{Keppens-JetMagField-2008}, the magnetic energy density $\rm U_b$ might be directly connected to the particles acceleration (energy gain -- bulk plasma heating) and jet structure collimation. 

Therefore, the argument put forward by \citet{Tavecchio1998} and \citet{Gao2011_LsyLIC} where the ratio between IC and Syn luminosities are directly related to the energy densities $\rm U_{ext}$ and $\rm U_b$ is based on the underling dynamic-mechanisms at work, i.e., the mechanisms responsible for particle acceleration and deceleration within the jet structure. In fact, given that synchrotron and external photons might undergo IC scattering,  $\rm L_{IC} \propto U'_{Ext} + U'_{Syn}$ should be more suitable for describing luminosity ratios in general,  and $\rm L_{IC} \propto U'_{Ext}$ might hold as the best approach to describe EC scenarios where $\rm U'_{Ext}$ is dominant with respect to $\rm U'_{Syn}$ (using ``$\prime$'' to  refer to jet rest-frame quantities):

\begin{equation}
\rm \frac{L_{IC}}{L_{Syn}}=   \frac{U'_{Ext}+U'_{Syn}}{U'_{B}} \ \ (a) \Rightarrow  \  \frac{U'_{Ext}}{U'_{B}}  \  \ (b)
\label{lumi-ratio}
\end{equation}

\begin{figure}[]
   \centering
    \includegraphics[width=1.0\linewidth]{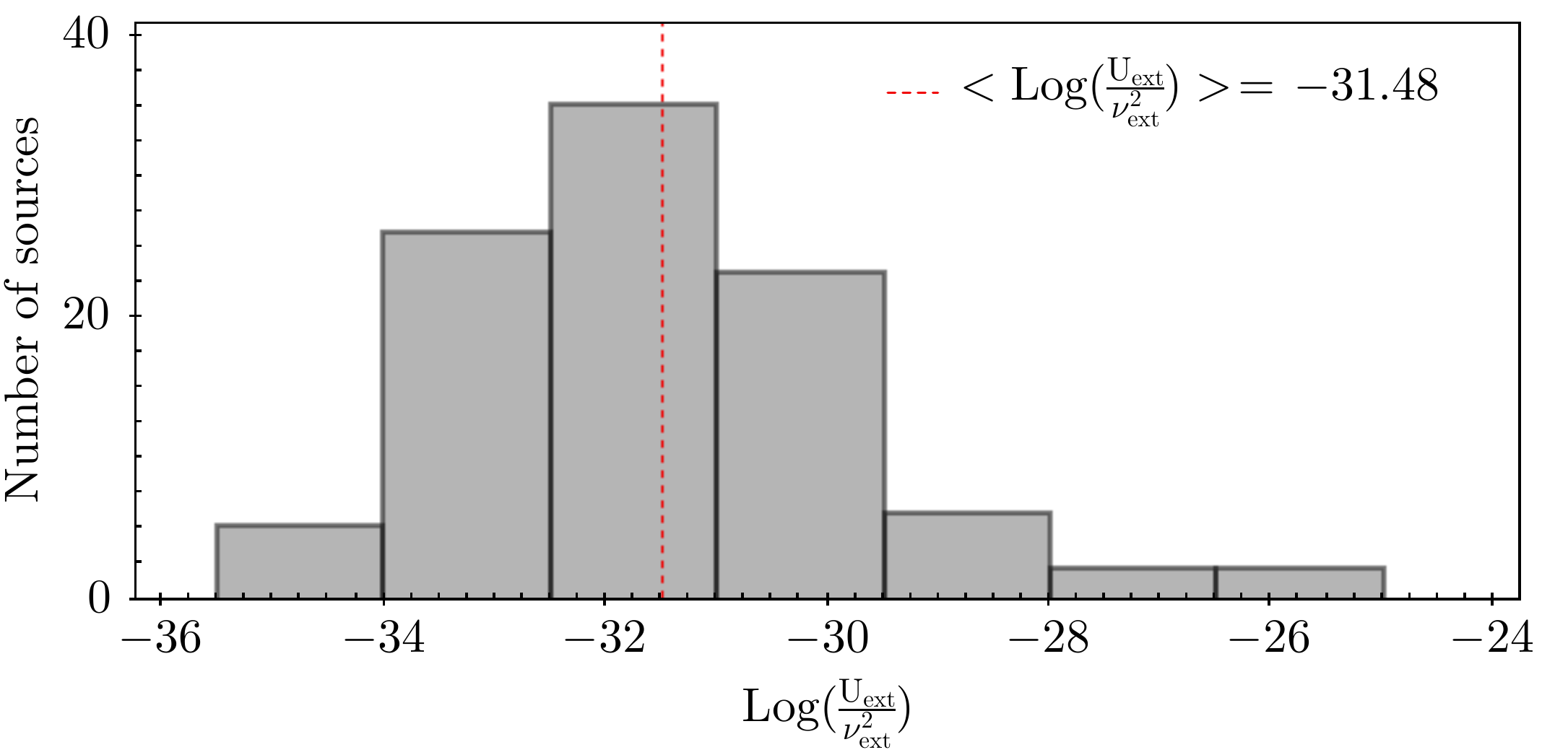}
     \caption{Distribution for the $\rm log(U_{ext}/\nu^2_{ext})$ parameter, as derived from eq. \ref{Gao} when considering measurements from Table \ref{tableRadioPlanck}. The mean value associated with $\rm log(U_{ext}/\nu^2_{ext})$ is highlighted with dashed red line.}
      \label{photon-density}
\end{figure} 

Also, we should note that the characteristic slope and tight correlation in the $\rm L_{IC}$ versus $ \rm L_{Syn}$ plane is in agreement with the CD distribution for LSP blazars \citep[as reported by ][]{paper1}, which is  Gaussian-like and peaks at log(CD) slightly higher than zero, at $\rm \approx0.17$. The fact that the slope associated with log(L$\rm _{IC}$) versus log(L$\rm _{Syn}$) is well established at $>$ 1.0 is probably related to the number of strong and fast flaring events in $\gamma$ rays which pushes the $\rm \langle L_{IC} \rangle $ to higher values when we integrate the observed flux from steady + flaring states over many years. In addition, it is telling us that the more powerful (luminous) blazars are the ones undergoing $\gamma$-ray flares more frequently. This could be a hint for the existence of an extra component apart from external and synchrotron photons that might be contributing to the IC bump during flaring events, especially for the most powerful (luminous) blazars. This is in agreement with the possibility of having hadronic or ultra-high-energy cosmic rays (UHECR) cascade components connected to the IC bump, just as considered by \cite{hadronic-model-PKS-2017}.

As discussed by \cite{Hu2017JetProp}, contributions from external photon fields (IR and UV, from accretion disk, BLR, and dust Torus) are relevant for describing the HE bump from blazar SEDs, and currently the major  difficulty is the lack of  precise knowledge about the AGN environment so that a multicomponent EC model can be fitted properly. In this scenario, it is hard to conclude the most relevant $\gamma$-ray emission site for individual sources, but as we describe here (from our population studies) the IR field tends to be more suitable to model the IC component of bright LSP blazars. Therefore, on average, a far-site emission  for MeV-GeV photons is favored, suggesting that an efficient acceleration mechanism might operate far from the core region, as mention by \cite{Sikora2009}.

From \cite{Tavecchio1998} and \cite{Gao2011_LsyLIC}, when using eq. \ref{lumi-ratio}.b, the external photon density U$_{ext}$ transforms to the jet comoving frame according to $\rm U'_{ext}= (17/12) \Gamma^2 U_{ext}$ as derived in  \cite{Ghisellini_Madau_1996}. Then, assuming the synchrotron peak $\rm \nu^{Syn}_{peak} = (4/3) \nu_L \gamma^2_{peak} \delta $ and $\rm \nu^{EC}_{peak} = (4/3) \nu^{ext}_{peak} \gamma^2_{peak} \Gamma \delta $, where $\rm \gamma_{peak}$ is the Lorentz factor for electrons emitting in the  Syn peak, $\rm \nu_L=eB/(2 \pi m_e c)$ is the Larmor frequency, and $\rm \nu^{ext}_{peak}$ is the peak frequency associated with the external photon field. From this \cite{Gao2011_LsyLIC} obtain

\begin{equation}
\rm \frac{L_{IC}}{L_{Syn}} \simeq \frac{17 e^2}{6 \pi m_e^2 c^2} \frac{U_{Ext}}{\nu^2_{ext}} \Big(  \frac{\nu^{EC}_{peak}}{\nu^S_{peak}} \Big)^2
\label{Gao}
\end{equation}


Using measured values for $\rm L_{Syn}$, $\rm L_{IC}$, and assuming $\rm \nu^{EC}_{peak} \equiv \nu^{IC}_{peak}$, with $ \rm \frac{17 e^2}{6 \pi m_e^2 c^2}=2.79 \times 10 ^{14}$ [$\rm \frac{cm^3}{erg.s^2}$], we infer the distribution of energy density $\rm (U_{ext}/\nu^2_{ext})$ associated with the external photon field at the AGN source frame (fig. \ref{photon-density}, which has mean value $\rm \langle log(U_{ext}/\nu^2_{ext})  \rangle = -32.53 \pm 0.17$ . Given the discussion from Sect. \ref{ECsection}, if we assume the external photon field to be dominant in IR, with $\rm \nu_{ext}=\nu_{IR}=3 \times 10 ^{13}$ Hz, the characteristic IR-photon energy density for LSP blazar under EC regime is $\rm \langle U_{ext} \rangle = 2.98 \times 10^{-5} $ erg/cm$^3$ at the AGN rest frame; The photon field seen by the jet (comoving jet frame) is then: $\rm U'_{ext}= (17/12) \Gamma^2 U_{ext} = 1.69 \times 10^{-2}$ erg/cm$^3$. Our estimate for $\rm \langle U'_{ext} \rangle$ is in good agreement with \cite{Breiding-2018-Sheath}, which assumes a far-site emission zone for $\gamma$-ray photons as a result of an IC upscattering of IR seed photons (originating from an illuminated molecular torus and assuming a spine-sheath geometry). We note that we follow the discussion from Sect. \ref{ECsection} and assume $\langle \Gamma \rangle = 20$, the bulk Lorentz factor associated with the relativistic outflow.  
 
\begin{figure}
   \centering
    \includegraphics[width=0.95\linewidth]{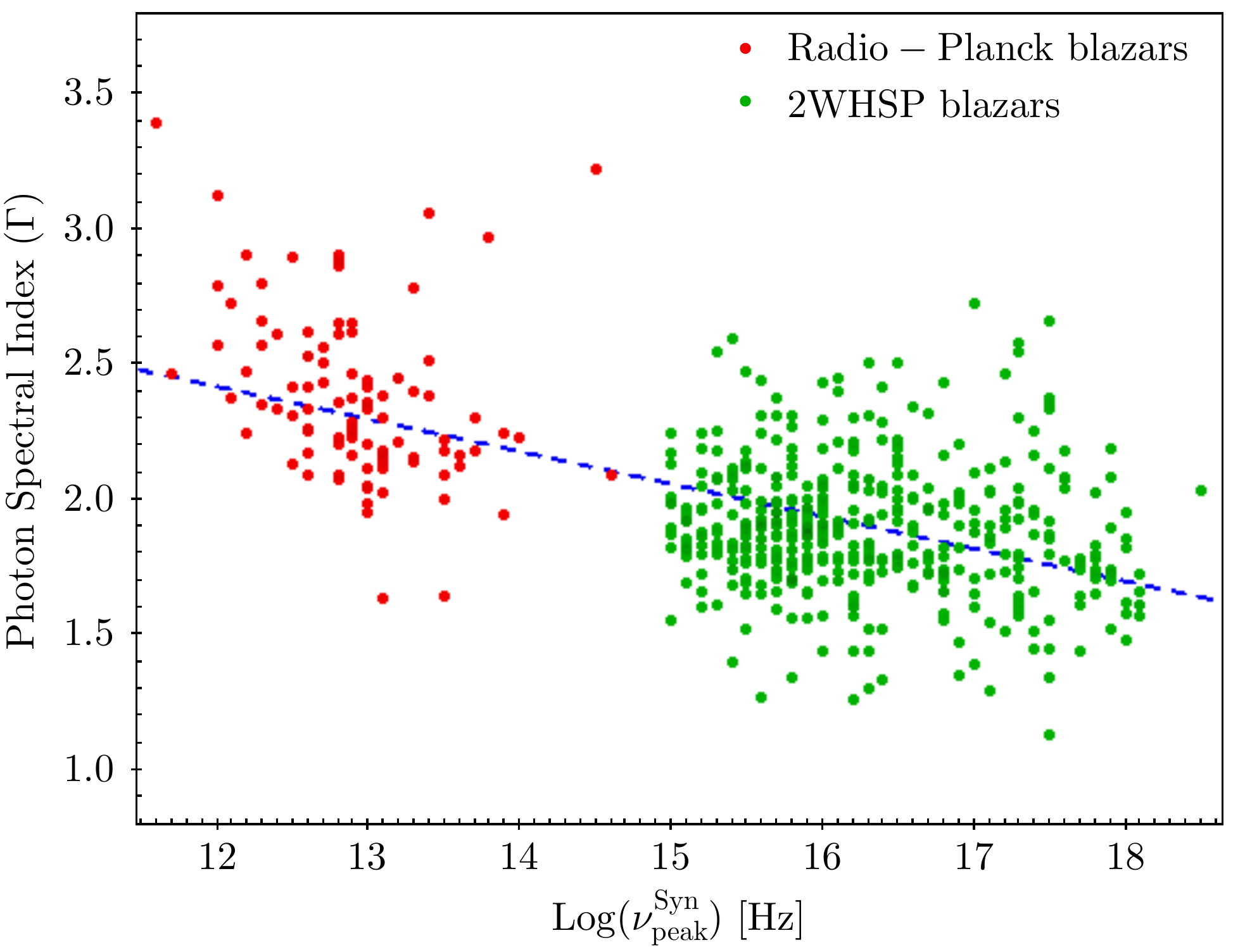}
     \caption{Gamma-ray photon spectral index ($\Gamma$) vs. the Syn peak $ \rm log(\nu_{peak}^{Syn}$) parameter: in red, the radio-Planck sources (LSP blazars); in green, the 2WHSP sources (HSP blazars); blue dashed line is the  linear fitting considering all sources in the $\Gamma$ vs. $ \rm log(\nu_{peak}^{Syn}$) plane.}
      \label{nusyngamma}
\end{figure}


\subsection{The $\gamma$-ray photon spectral index versus the synchrotron peak frequency}

In Fig. \ref{nusyngamma}  we show the correlation between the $\gamma$-ray photon spectral index and the logarithm  of the Syn peak frequency log($\nu_{peak}^{Syn}$), considering all the 99 radio-Planck sources that had available data in the MeV to GeV band. To expand the description beyond LSP sources and extend the test to higher $\nu^{sync}_{peak}$ values, in this same plot we add the 2WHSP sources \citep[][]{2WHSP} which is a highly confident sample of high synchrotron peak (HSP) blazars. A linear fitting in the $\Gamma$ versus log($\nu_{peak}^{Syn}$) plane reveals a clear negative trend,
\begin{equation}
\rm \langle \Gamma \rangle=-0.119 \times log( \langle \nu_{peak}^{Syn} \rangle )+3.85
\label{corr1}
,\end{equation}
showing,  on average, that increasing synchrotron peak frequency is related to the hardening of the $\gamma$-ray spectrum  in the  0.1 to 500 GeV band, as also reported by \citet{3FGL} and \citet{1WHSP}. This is usually explained as a consequence of the fixed observational energy window from \textit{Fermi}-LAT ($\sim$100 MeV up to 500 GeV) which probes different regions of the IC bump: after its peak (soft spectra with decaying power $\rm P_{(\nu)}$) in case of LSPs, and before its peak (hard spectra  with increasing power $\rm P_{(\nu)}$) in the case of HSPs. This is usually taken as observational evidence that $\nu^{IC}_{peak}$ is moving to higher energies according to $\nu^{Syn}_{peak}$.

\begin{figure}[]
   \centering
    \includegraphics[width=0.8\linewidth]{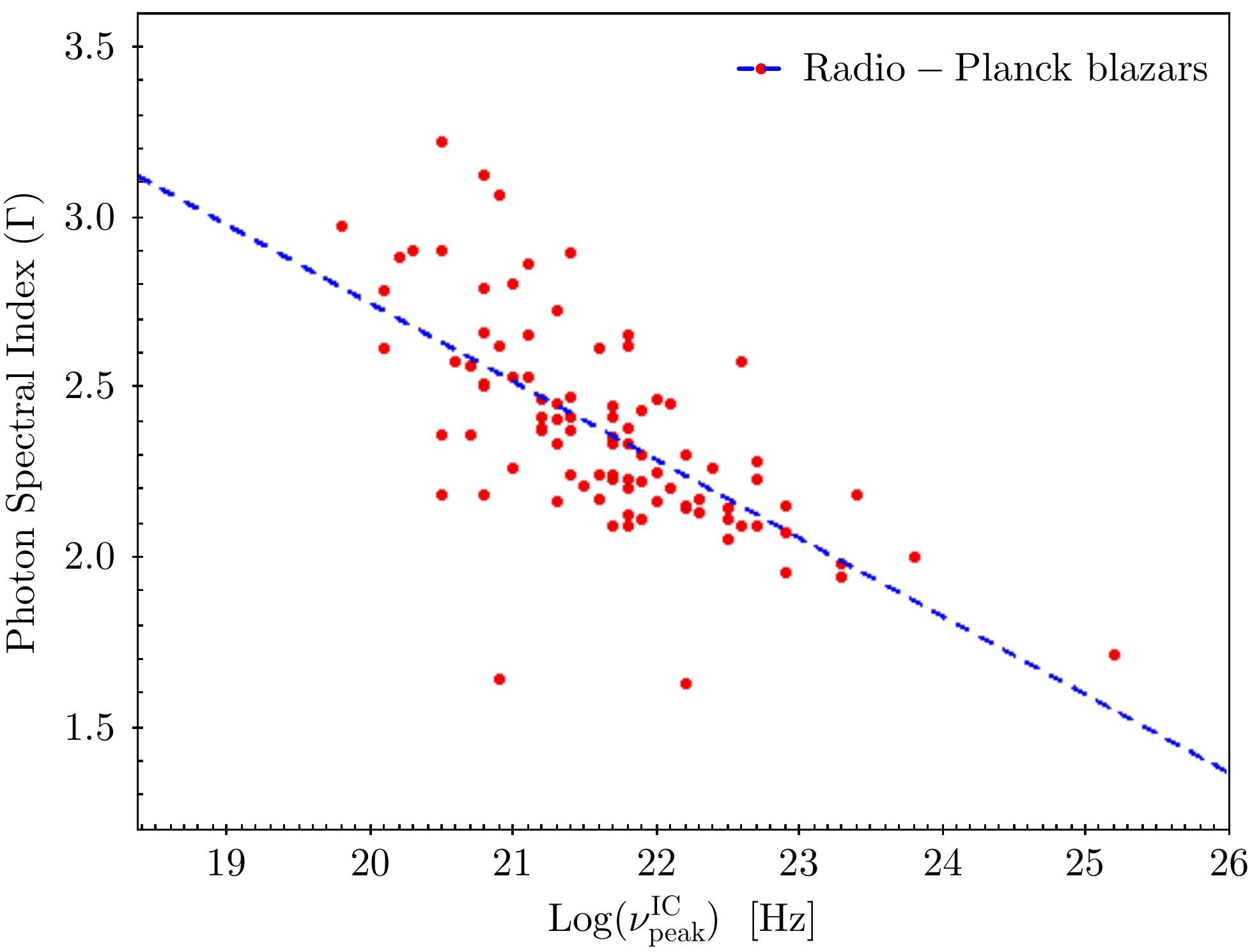}
     \caption{Gamma-ray photon spectral index ($\Gamma$) vs. the inverse Compton peak $\rm log(\nu_{peak}^{IC}$) parameter. In red, the radio-Planck sources with good estimates for the IC peak; the blue dashed line is the linear fitting for the data.}
      \label{nuicgamma}
\end{figure}

The connection between $\nu^{IC}_{peak}$ and $\nu^{Syn}_{peak}$ is actually very hard to probe directly, simply because we have limited data to describe the IC bump in the case of HSP blazars. For HSPs, the IC bump extends farther than the \textit{Fermi}-LAT main sensitivity window, and despite the many detections with ground-based very high-energy observatories (VHE, at E\,$>$\,100\,GeV), the absorption of VHE photons due to scattering with low-energy extragalactic background light (EBL) hinders the description of the IC peak. 

If we consider only the complete radio-Planck sample of LSP blazars \citep{paper1}, a scatter plot with $\rm log(\nu^{Syn}_{peak})$ versus $\rm log(\nu^{IC}_{peak})$ shows no clear correlation. A complete sample of blazars spanning a wider  range in $\rm log(\nu_{peak})$ space might be needed to better probe the $\rm \nu^{Syn}_{peak}$ to $\rm \nu^{IC}_{peak}$ connection. In Fig. \ref{nuicgamma} we plot the $\gamma$-ray photon spectral index against the $ \rm log(\nu_{peak}^{IC}$), this time only for the radio-Planck sources with good estimates for the IC parameter (cases with the ? flag in Table \ref{tableRadioPlanck} were eliminated). Even if we try to use a complete sample of HSP blazars, there is no good estimate of $\nu^{IC}_{peak}$ for all sources, and therefore we do not consider HSPs for this plot. A linear fitting in the $\Gamma$ versus $ \rm log(\nu_{peak}^{IC}$) gives  

\begin{equation}
\rm \langle \Gamma \rangle=-0.229 \times log( \langle \nu_{peak}^{IC} \rangle )+7.34
\label{corr2}
\end{equation}

Both correlations, as in eq. \ref{corr1} and eq. \ref{corr2}, tell us that LSP blazars are associated with the steepest $\gamma$-ray sources in the 0.1--500 GeV band, with an IC peak located around the MeV band. Faint point-like sources of this kind are difficult to detect with \Fermi-LAT, especially in regions close to the galactic disk $|b|<10^\circ$ where the MeV diffuse component is dominant. In pure leptonic SSC and EC scenarios, a correlation between spectral parameters derived from the Syn and IC components is expected \citep{Giommi-Blazar-fog-2012,Giommi-gamma-case-2013} given that both components depend directly on the jet's relativistic electrons producing synchrotron radiation and acting for the up-scattering of low-energy photons to $\gamma$ rays.


\section{Conclusions}

We evaluate the jet's Lorentz factor $\gamma_{peak}$ and B$\delta$ parameters for LSP blazars in the radio-Planck sample, assuming at first a simple single-zone SSC model. In this case, we show that B$\delta$ is probably underestimated owing to overestimated $\gamma^{SSC}_{peak}$ values; therefore, a SSC model can hardly describe the SED observed for LSP blazars. 

We studied the Tramacere plane $\rm log(\gamma^{SSC}_{peak})$ versus $\rm log(\nu^{Syn}_{peak})$ to show that most sources in the radio-Planck sample are above the limits  associated with a dominant SSC regime. In fact, they populate a region that is characteristic of the EC regime under TH scattering, spreading along a parameter-space that is attributed to external photon fields ranging from IR to UV. 

Assuming an EC model, we reevaluate the $\gamma^{EC}_{peak}$ and B$\delta$ parameters for LSP blazars. We assume two different external photon fields,  one dominated by UV photons (consistent with BLR emission) and another dominated by IR photons (consistent with dust torus emission, and MC emission in spine-sheath geometry). We conclude that on average an IR field is probably more suitable, resulting in distributions with the corresponding mean values $\rm \langle log(B\delta) \rangle \approx 0.99$ and $\rm \langle log(\gamma^{EC}_{peak}) \rangle \approx 2.80 $ consistent with expectations from \citet{Kang2014_Lorentzfactor} and \citet{Gang2013_LocationGamma}. This hints to a $\gamma$-ray emission region which is out of the BLR domain, far from the BH, at a distance $\gg$ 0.1 pc. Moreover, it demands the jet structure to be a very narrow opening (or with substructures) to reconcile with the short timescale variability observed in the GeV-TeV band \citep{Agudo2011}. 
We calculate the photon energy density associated with the external field at the jet comoving frame to be $\rm U'_{ext} = 1.69 \times 10^{-2}$ erg/cm$^3$, finding good agreement with \cite{Breiding-2018-Sheath}.


We calculate the luminosity associated with the peak-power for both Syn and IC components, and plot $\rm log(L_{IC})$ versus $ \rm log(L_{Syn})$ in what we called {``the luminosity plane''}. There we show a tight correlation spanning seven orders of magnitude in luminosity, with slope slightly larger than one, which is probably related to a nearly constant ratio of the energy density associated with external + synchrotron photon fields to the magnetic energy density, ($\rm U^\prime_{Ext}+U^\prime_{Syn})/U^\prime_B$, implying a balance between the  particle's acceleration and deceleration mechanisms in the jet. In fact, the slope we measure in the luminosity plane is larger than 1.0 and could be induced by the $\gamma$-ray flaring activity, which is proving to be more relevant for the most luminous (powerful and extreme) sources.   

We probe the correlation between the $\gamma$-ray photon spectral index (0.1--500 GeV band) with both $\rm \nu^{Syn}_{peak}$ and $\rm \nu^{IC}_{peak}$ parameters, showing a trend of hardening $\Gamma$ for increasing $\nu^{Syn}_{peak}$ and $\rm \nu^{IC}_{peak}$, noting that LSP blazars are characterized by steep $\gamma$-ray spectrum in the 0.1--500\,GeV band, which hinders the detection of faint LSP sources with \Fermi-LAT.    

\begin{acknowledgements}

During this work, BA was supported by the Brazilian Scientific Program Ci\^{e}ncias sem Fronteiras - Cnpq, and later by S\~ao Paulo Research Foundation (FAPESP) with grant n. 2017/00517-4. YLC is supported by the Government of the Republic of China (Taiwan). We would like to thank Prof. Paolo Giommi and Prof. Gianluca Polenta for their comments during the preparation of this work, Prof. Marcelo M. Guzzo and Prof. Orlando L.G. Peres for the full support granting the author's partnership with FAPESP. We thank SSDC, the Space Science Data Center from the Agenzia Spaziale Italiana; University La Sapienza of Rome, Department of Physics; And State University of Campinas - Unicamp, IFGW Department of Physics for hosting the authors. We make use of archival data and bibliographic information obtained from the NASA-IPAC Extragalactic Database (NED), and data and software facilities from the SSDC (\url{www.ssdc.asi.it}). 

\end{acknowledgements}

\bibliographystyle{aa}
\bibliography{Fermipaper}


\begin{longtable}
{>{\raggedright\arraybackslash}p{2.8cm}>{\centering\arraybackslash}p{1.0cm}|>{\centering\arraybackslash}p{1.0cm}>{\centering\arraybackslash}p{1.0cm}>{\centering\arraybackslash}p{1.0cm}>{\centering\arraybackslash}p{1.0cm}>{\centering\arraybackslash}p{1.0cm}>{\centering\arraybackslash}p{1.0cm}>
{\centering\arraybackslash}p{1.0cm}>
{\centering\arraybackslash}p{1.0cm}>
{\centering\arraybackslash}p{1.0cm}}
\caption{Here we lits all 104 sources used for our studies. Column 5BZcat shows the blazar name according to \cite{5BZcat} where BZQ stands for Flat Spectrum Radio Quasars, BZB for BL Lacs and BZU for still undefined-class blazars. Column z corresponds to the redshift as reported in the 5BZcat \cite{5BZcat}, and from the NASA-IPAC Extragalactic Database (NED). We list the fitting parameters characterizing the peak power for the synchrotron component log($\rm \nu_{Syn}$) [Hz]; log($\rm \nu$f$^{Syn}_{\nu}$) [erg/cm$^2$/s] and inverse Compton component log($\rm \nu_{IC}$) [Hz]; log($\rm \nu$f$^{IC}_{\nu}$) [erg/cm$^2$/s] as measured from the mean SED when considering all available data \citep{paper1}. The parameter log($\rm \gamma^{SSC}_{peak}$) corresponds to the Lorentz factor associated to relativistic electrons at the synchrotron peak-frequency when considering a pure SSC model (* in many cases leading to overestimate values). The parameter log($\rm \gamma^{EC}_{IR/UV}$) corresponds to the Lorentz factor when assuming an EC scenario with dominant IR seed photons from the dust and MC torus (denoted with IR suffix), and UV seed photons from accretion disk and BLR region (denoted with UV suffix). The log(B$\rm \delta_{IR}$) and log(B$\rm \delta_{UV}$) columns list the corresponding values for B$\rm \delta$ when assuming a dominant IR and UV external photon fields, respectively, with B given in {\it Gauss} and $\rm \delta$ representing the beaming factor. We would like to note that \cite{paper1} also list R.A., Dec. (J2000) and radio counterparts for each source according to NVSS catalog \citep{NVSS}.}\\
\hline\hline
   &  & log& log& log &  log& log& log& log& log& log\\
5BZcat\,J   & z   & $\nu_{Syn}$ & $\nu$f$^{Syn}_{\nu}$ & $\nu_{IC}$&  $\nu$f$^{IC}_{\nu}$& $\gamma^{SSC}_{peak}$ & $\gamma^{EC}_{IR}$ & B$\delta_{IR}$&$\gamma^{EC}_{UV}$ & B$\delta_{UV}$\\
\hline
\endfirsthead
\caption{continued.}\\
\hline\hline
   &  & log& log& log &  log& log& log& log& log& log\\
5BZcat\,J   & z   & $\nu_{Syn}$ & $\nu$f$^{Syn}_{\nu}$ & $\nu_{IC}$&  $\nu$f$^{IC}_{\nu}$& $\gamma^{SSC}_{peak}$ & $\gamma^{EC}_{IR}$ & B$\delta_{IR}$&$\gamma^{EC}_{UV}$ & B$\delta_{UV}$\\
\hline
\endhead
\hline
\endfoot 
  5BZBJ0050$-$0929    & $...$   &  14.6 &  $-$11.0 &  22.7 &  $-$11.1   & 3.98 &  3.24 &  1.59 &  2.33 &  2.51   \\    
  5BZBJ0238+1636    &  0.94   &  13.0 &  $-$10.9 &  22.5 &  $-$10.4   & 4.68 &  3.29 &  0.19 &  2.37 &  1.11   \\    
  5BZBJ0449+1121    &  2.153  &  12.9 &  $-$11.5 &  21.8 &  $-$10.8   & 4.38 &  3.04 &  0.79 &  2.13 &  1.71   \\    
  5BZBJ0721+7120    &  $...$    &  13.9 &  $-$10.3 &  23.3 &  $-$10.5   & 4.63 &  3.54 &  0.29 &  2.63 &  1.21   \\    
  5BZBJ0738+1742    &  0.424  &  13.5 &  $-$10.6 &  23.8 &  $-$11.0   & 5.08 &  3.87 & $-$0.60 &  2.96 &  0.31   \\    
  5BZBJ0757+0956    &  0.266  &  13.7 &  $-$10.8 &  20.8 &  $-$11.1   & 3.48 &  2.34 &  2.59 &  1.43 &  3.51   \\    
  5BZBJ0825+0309    &  0.506  &  13.1 &  $-$11.1 &  21.5(?) &  $-$11.6   & 4.13 &  2.73 &  1.29 &  1.82 &  2.21   \\    
  5BZBJ0854+2006    &  0.306  &  13.6 &  $-$10.4 &  21.8 &  $-$10.8   & 4.03 &  2.85 &  1.49 &  1.94 &  2.41   \\    
  5BZBJ0958+6533    &  0.367  &  13.4 &  $-$11.0 &  21.2 &  $-$11.3   & 3.83 &  2.56 &  1.89 &  1.65 &  2.81   \\    
  5BZBJ1419+5423    &  0.153  &  13.7 &  -10.6 &  21.9 &  $-$11.5   & 4.03 &  2.87 &  1.49 &  1.96 &  2.41   \\    
  5BZBJ1653+3945    &  0.033  &  17.8 &  $-$10.2 &  25.2 &  $-$10.6   & 3.63 &  4.50 &  2.29 &  3.59 &  3.21   \\    
  5BZBJ1800+7828    &  0.68   &  13.5 &  $-$10.7 &  21.9 &  $-$10.9   & 4.13 &  2.96 &  1.29 &  2.04 &  2.21   \\    
  5BZBJ1806+6949    &  0.046  &  14.0 &  $-$10.6 &  21.7 &  $-$11.0   & 3.78 &  2.75 &  1.99 &  1.84 &  2.91   \\    
  5BZBJ1824+5651    &  0.663  &  13.2 &  $-$11.3 &  22.1 &  $-$11.0   & 4.38 &  3.05 &  0.79 &  2.14 &  1.71   \\    
  5BZBJ2005+7752    &  0.342  &  13.2 &  $-$11.2 &  21.5 &  $-$11.3   & 4.08 &  2.71 &  1.39 &  1.79 &  2.31   \\    
  5BZBJ2134$-$0153    &  1.283  &  12.8 &  $-$11.3 &  21.8 &  $-$11.5   & 4.43 &  2.97 &  0.69 &  2.06 &  1.61   \\    
  5BZBJ2202+4216    &  0.069  &  13.6 &  $-$10.0 &  21.3 &  $-$10.3   & 3.78 &  2.56 &  1.99 &  1.65 &  2.91   \\    
  5BZQJ0006$-$0623    &  0.347  &  13.0 &  $-$11.1 &  20.3 &  $-$11.9   & 3.58 &  2.11 &  2.39 &  1.20 &  3.31   \\    
  5BZQJ0010+1058    &  0.089  &  14.5 &  $-$10.7 &  20.5 &  $-$10.8   & 2.93 &  2.16 &  3.69 &  1.25 &  4.61   \\    
  5BZQJ0108+0135    &  2.099  &  12.9 &  $-$11.2 &  22.4 &  $-$10.6   & 4.68 &  3.34 &  0.19 &  2.43 &  1.11   \\    
  5BZQJ0125$-$0005    &  1.077  &  12.8 &  $-$11.6 &  20.3 &  $-$11.6   & 3.68 &  2.20 &  2.19 &  1.29 &  3.11   \\    
  5BZQJ0136+4751    &  0.859  &  13.3 &  $-$10.9 &  22.2 &  $-$10.7   & 4.38 &  3.13 &  0.79 &  2.22 &  1.71   \\    
  5BZQJ0152+2207    &  1.32   &  12.9 &  $-$11.5 &  21.8 &  $-$11.5   & 4.38 &  2.98 &  0.79 &  2.06 &  1.71   \\    
  5BZQJ0217+7349    &  2.367  &  12.2 &  $-$11.6 &  20.5 &  $-$10.4   & 4.08 &  2.41 &  1.39 &  1.49 &  2.31   \\    
  5BZQJ0228+6721    &  0.523  &  12.8 &  $-$11.2 &  21.1 &  $-$11.4   & 4.08 &  2.53 &  1.39 &  1.62 &  2.31   \\    
  5BZQJ0237+2848    &  1.206  &  12.9 &  $-$11.2 &  22.0 &  $-$10.7   & 4.48 &  3.06 &  0.59 &  2.15 &  1.51   \\    
  5BZQJ0309+1029    &  0.863  &  12.8 &  $-$11.2 &  21.7 &  $-$11.2   & 4.38 &  2.88 &  0.79 &  1.97 &  1.71   \\    
  5BZQJ0336+3218    &  1.259  &  12.8 &  $-$11.3 &  20.2 &  $-$10.5   & 3.63 &  2.17 &  2.29 &  1.26 &  3.21   \\    
  5BZQJ0339$-$0146    &  0.805  &  12.6 &  $-$11.3 &  22.0 &  $-$11.1   & 4.63 &  3.02 &  0.29 &  2.11 &  1.21   \\    
  5BZQJ0359+5057    &  1.512  &  12.1 &  $-$10.7 &  21.3 &  $-$10.4   & 4.53 &  2.74 &  0.49 &  1.83 &  1.41   \\    
  5BZQJ0423$-$0120    &  0.916  &  13.0 &  $-$10.6 &  22.1 &  $-$10.7   & 4.48 &  3.08 &  0.59 &  2.17 &  1.51   \\    
  5BZQJ0501$-$0159    &  2.291  &  13.0 &  $-$11.4 &  21.4 &  $-$11.1   & 4.13 &  2.85 &  1.29 &  1.94 &  2.21   \\    
  5BZQJ0510+1800    &  0.416  &  13.3 &  $-$11.3 &  21.3 &  $-$11.1   & 3.93 &  2.62 &  1.69 &  1.71 &  2.61   \\    
  5BZQJ0530+1331    &  2.07   &  12.2 &  $-$11.5 &  21.4 &  $-$10.7   & 4.53 &  2.84 &  0.49 &  1.92 &  1.41   \\    
  5BZQJ0555+3948    &  2.365  &  12.0 &  $-$11.7 &  20.8 &  $-$10.9   & 4.33 &  2.56 &  0.89 &  1.64 &  1.81   \\    
  5BZQJ0607$-$0834    &  0.87   &  12.1 &  $-$11.5 &  21.4 &  $-$11.0   & 4.58 &  2.73 &  0.39 &  1.82 &  1.31   \\    
  5BZQJ0646+4451    &  3.396  &  11.6 &  $-$11.8 &  21.8(?) &  $-$11.5   & 5.03 &  3.11 & $-$0.50 &  2.20 &  0.41   \\    
  5BZQJ0739+0137    &  0.189  &  13.9 &  $-$10.9 &  21.6 &  $-$10.8   & 3.78 &  2.73 &  1.99 &  1.82 &  2.91   \\    
  5BZQJ0750+1231    &  0.889  &  12.6 &  $-$11.1 &  21.2 &  $-$11.2   & 4.23 &  2.63 &  1.09 &  1.72 &  2.01   \\    
  5BZQJ0808+4950    &  1.432  &  12.0 &  $-$12.2 &  20.6 &  $-$11.7   & 4.23 &  2.39 &  1.09 &  1.47 &  2.01   \\    
  5BZQJ0808$-$0751    &  1.837  &  13.0 &  $-$11.1 &  22.9 &  $-$10.7   & 4.88 &  3.57 & $-$0.20 &  2.66 &  0.71   \\    
  5BZQJ0830+2410    &  0.939  &  12.6 &  $-$11.1 &  21.8 &  $-$11.0   & 4.53 &  2.94 &  0.49 &  2.02 &  1.41   \\    
  5BZQJ0841+7053    &  2.218  &  12.4 &  $-$11.3 &  20.1 &  $-$10.2   & 3.78 &  2.20 &  1.99 &  1.28 &  2.91   \\    
  5BZQJ0920+4441    &  2.19   &  12.5 &  $-$11.2 &  22.3 &  $-$10.6   & 4.83 &  3.29 & $-$0.10 &  2.38 &  0.81   \\    
  5BZQJ0927+3902    &  0.695  &  12.1 &  $-$11.2 &  $...$  &  $...$     & $-$6.1 &  $-$7.9 &  21.7 & $-$8.89 &  22.7   \\    
  5BZQJ0948+4039    &  1.249  &  12.3 &  $-$11.8 &  20.8 &  $-$11.2   & 4.18 &  2.47 &  1.19 &  1.56 &  2.11   \\    
  5BZQJ0956+2515    &  0.712  &  12.7 &  $-$11.4 &  21.9 &  $-$11.4   & 4.53 &  2.96 &  0.49 &  2.05 &  1.41   \\    
  5BZQJ1038+0512    &  0.473  &  12.0 &  $-$11.8 &  20.8 &  $-$12.1   & 4.33 &  2.38 &  0.89 &  1.47 &  1.81   \\    
  5BZQJ1043+2408    &  0.56   &  12.9 &  $-$11.5 &  21.7 &  $-$11.6   & 4.33 &  2.84 &  0.89 &  1.93 &  1.81   \\    
  5BZQJ1130+3815    &  1.733  &  12.3 &  $-$11.7 &  22.6 &  $-$11.6   & 5.08 &  3.41 & -0.60 &  2.50 &  0.31   \\    
  5BZQJ1153+4931    &  0.334  &  12.9 &  $-$10.9 &  21.2 &  $-$11.1   & 4.08 &  2.56 &  1.39 &  1.64 &  2.31   \\    
  5BZQJ1153+8058    &  1.25   &  12.6 &  $-$12.0 &  21.1 &  $-$11.9   & 4.18 &  2.62 &  1.19 &  1.71 &  2.11   \\    
  5BZQJ1159+2914    &  0.729  &  13.5 &  $-$11.1 &  22.6 &  $-$10.8   & 4.48 &  3.31 &  0.59 &  2.40 &  1.51   \\    
  5BZQJ1222+0413    &  0.966  &  12.7 &  $-$11.2 &  20.8 &  $-$10.7   & 3.98 &  2.44 &  1.59 &  1.53 &  2.51   \\    
  5BZQJ1224+2122    &  0.434  &  13.1 &  $-$10.7 &  23.0 &  $-$10.0   & 4.88 &  3.47 & $-$0.20 &  2.56 &  0.71   \\    
  5BZQJ1229+0203    &  0.158  &  13.4 &  $-$10.0 &  20.8 &  $-$9.54   & 3.63 &  2.32 &  2.29 &  1.41 &  3.21   \\    
  5BZQJ1256$-$0547    &  0.536  &  12.8 &  $-$10.0 &  22.7 &  $-$10.0   & 4.88 &  3.34 & $-$0.20 &  2.42 &  0.71   \\    
  5BZQJ1327+2210    &  1.398  &  12.6 &  $-$11.7 &  21.7 &  $-$11.0   & 4.48 &  2.93 &  0.59 &  2.02 &  1.51   \\    
  5BZQJ1504+1029    &  1.839  &  12.8 &  $-$11.6 &  22.9 &  $-$10.0   & 4.98 &  3.57 & $-$0.40 &  2.66 &  0.51   \\    
  5BZQJ1512$-$0905    &  0.36   &  13.1 &  $-$10.9 &  22.2 &  $-$9.91   & 4.48 &  3.06 &  0.59 &  2.15 &  1.51   \\    
  5BZQJ1549+0237    &  0.414  &  12.9 &  $-$11.2 &  22.0 &  $-$11.1   & 4.48 &  2.97 &  0.59 &  2.06 &  1.51   \\    
  5BZQJ1550+0527    &  1.422  &  13.0 &  $-$11.4 &  21.8 &  $-$11.4   & 4.33 &  2.99 &  0.89 &  2.07 &  1.81   \\    
  5BZQJ1608+1029    &  1.226  &  12.8 &  $-$11.3 &  21.6 &  $-$11.0   & 4.33 &  2.87 &  0.89 &  1.95 &  1.81   \\    
  5BZQJ1613+3412    &  1.397  &  12.3 &  $-$11.5 &  21.7 &  $-$11.6   & 4.63 &  2.93 &  0.29 &  2.02 &  1.21   \\    
  5BZQJ1635+3808    &  1.814  &  12.5 &  $-$11.1 &  21.7 &  $-$10.3   & 4.53 &  2.97 &  0.49 &  2.06 &  1.41   \\    
  5BZQJ1640+3946    &  1.66   &  12.9 &  $-$11.9 &  22.7 &  $-$10.8   & 4.83 &  3.46 & $-$0.10 &  2.54 &  0.81   \\    
  5BZQJ1642+3948    &  0.593  &  13.0 &  $-$10.6 &  21.7 &  $-$10.7   & 4.28 &  2.84 &  0.99 &  1.93 &  1.91   \\    
  5BZQJ1642+6856    &  0.751  &  12.5 &  $-$11.6 &  20.1(?) &  $-$12.3   & 3.73 &  2.06 &  2.09 &  1.15 &  3.01   \\    
  5BZQJ1740+5211    &  1.381  &  13.2 &  $-$11.3 &  21.3 &  $-$10.8   & 3.98 &  2.73 &  1.59 &  1.82 &  2.51   \\    
  5BZQJ1743$-$0350    &  1.057  &  12.6 &  $-$11.3 &  21.0 &  $-$11.2   & 4.13 &  2.55 &  1.29 &  1.64 &  2.21   \\    
  5BZQJ1849+6705    &  0.657  &  13.1 &  $-$10.9 &  22.5 &  $-$10.6   & 4.63 &  3.25 &  0.29 &  2.34 &  1.21   \\    
  5BZQJ1927+7358    &  0.302  &  13.1 &  $-$10.8 &  $...$  &  $...$     & $-$6.6 &  $-$8.0 &  22.7 & $-$8.95 &  23.7   \\    
  5BZQJ1955+5131    &  1.214  &  12.7 &  $-$11.7 &  20.7 &  $-$11.4   & 3.93 &  2.42 &  1.69 &  1.50 &  2.61   \\    
  5BZQJ2007+4029    &  1.736  &  12.2 &  $-$11.6 &  $...$  &  $...$     & $-$6.1 &  $-$7.8 &  21.8 & $-$8.79 &  22.8   \\    
  5BZQJ2022+6136    &  0.228  &  12.9 &  $-$11.2 &  $...$  &  $...$     & $-$6.5 &  $-$8.0 &  22.5 & $-$8.96 &  23.5   \\    
  5BZQJ2038+5119    &  1.686  &  12.5 &  $-$11.5 &  21.4 &  $-$11.0   & 4.38 &  2.81 &  0.79 &  1.90 &  1.71   \\    
  5BZQJ2123+0535    &  1.941  &  12.6 &  $-$11.7 &  21.6 &  $-$11.8   & 4.43 &  2.93 &  0.69 &  2.02 &  1.61   \\    
  5BZQJ2136+0041    &  1.941  &  11.7 &  $-$11.6 &  21.2 &  $-$11.2   & 4.68 &  2.73 &  0.19 &  1.82 &  1.11   \\    
  5BZQJ2139+1423    &  2.427  &  12.2 &  $-$11.7 &  $...$  &  $...$     & $-$6.1 &  $-$7.85&  21.8 & $-$8.74 &  22.8   \\    
  5BZQJ2148+0657    &  0.999  &  12.5 &  $-$10.8 &  20.5 &  $-$11.0   & 3.93 &  2.29 &  1.69 &  1.38 &  2.61   \\    
  5BZQJ2203+1725    &  1.076  &  13.3 &  $-$10.9 &  22.9 &  $-$10.9   & 4.73 &  3.50 &  0.09 &  2.59 &  1.01   \\    
  5BZQJ2203+3145    &  0.295  &  13.4 &  $-$10.9 &  20.9 &  $-$11.1   & 3.68 &  2.40 &  2.19 &  1.49 &  3.11   \\    
  5BZQJ2218$-$0335    &  0.901  &  12.3 &  $-$11.3 &  21.0 &  $-$11.7   & 4.28 &  2.53 &  0.99 &  1.62 &  1.91   \\    
  5BZQJ2225$-$0457    &  1.404  &  13.0 &  $-$10.8 &  21.7 &  $-$10.9   & 4.28 &  2.93 &  0.99 &  2.02 &  1.91   \\    
  5BZQJ2229$-$0832    &  1.56   &  13.1 &  $-$11.2 &  21.8 &  $-$10.5   & 4.28 &  3.00 &  0.99 &  2.09 &  1.91   \\    
  5BZQJ2232+1143    &  1.037  &  12.4 &  $-$11.2 &  21.3 &  $-$10.5   & 4.38 &  2.70 &  0.79 &  1.79 &  1.71   \\    
  5BZQJ2236+2826    &  0.79   &  13.0 &  $-$11.2 &  22.5 &  $-$10.9   & 4.68 &  3.27 &  0.19 &  2.36 &  1.11   \\    
  5BZQJ2253+1608    &  0.859  &  13.1 &  $-$10.0 &  22.2 &  $-$9.2    & 4.48 &  3.13 &  0.59 &  2.22 &  1.51   \\    
  5BZQJ2354+4553    &  1.992  &  12.2 &  $-$11.9 &  21.4 &  $-$11.7   & 4.53 &  2.83 &  0.49 &  1.92 &  1.41   \\    
  5BZQJ2356+8152    &  1.344  &  12.8 &  $-$11.8 &  21.1 &  $-$11.5   & 4.08 &  2.63 &  1.39 &  1.72 &  2.31   \\    
  5BZUJ0102+5824    &  0.644  &  12.6 &  $-$11.1 &  21.8 &  $-$10.9   & 4.53 &  2.90 &  0.49 &  1.99 &  1.41   \\    
  5BZUJ0204+1514    &  0.833  &  12.6 &  $-$11.6 &  21.0 &  $-$11.2   & 4.13 &  2.52 &  1.29 &  1.61 &  2.21   \\    
  5BZUJ0241$-$0815    &  0.005  &  13.5 &  $-$10.1 &  20.9(?) &  $-$10.6   & 3.63 &  2.34 &  2.29 &  1.43 &  3.21   \\    
  5BZUJ0319+4130    &  0.018  &  13.0 &  $-$10.4 &  23.3 &  $-$10.4   & 5.08 &  3.55 & $-$0.60 &  2.63 &  0.31   \\    
  5BZUJ0433+0521    &  0.033  &  13.8 &  $-$10.2 &  19.8 &  $-$10.2   & 2.93 &  1.80 &  3.69 &  0.89 &  4.61   \\    
  5BZUJ0725$-$0054    &  0.128  &  13.5 &  $-$11.0 &  20.5 &  $-$11.2   & 3.43 &  2.17 &  2.69 &  1.26 &  3.61   \\    
  5BZUJ0909+4253    &  0.67   &  12.9 &  $-$11.5 &  20.9 &  $-$11.7   & 3.93 &  2.45 &  1.69 &  1.54 &  2.61   \\    
  5BZUJ1058+0133    &  0.89   &  13.1 &  $-$10.9 &  22.3 &  $-$10.8   & 4.53 &  3.18 &  0.49 &  2.27 &  1.41   \\    
  5BZUJ1310+3220    &  0.997  &  13.1 &  $-$10.8 &  22.2 &  $-$10.8   & 4.48 &  3.14 &  0.59 &  2.23 &  1.51   \\    
  5BZUJ1415+1320    &  0.247  &  12.8 &  $-$11.0 &  20.5 &  $-$11.2   & 3.78 &  2.19 &  1.99 &  1.28 &  2.91   \\    
  5BZUJ1751+0939    &  0.322  &  13.1 &  $-$10.8 &  21.9 &  $-$10.8   & 4.33 &  2.90 &  0.89 &  1.99 &  1.81   \\    
  5BZUJ1829+4844    &  0.695  &  13.0 &  $-$11.3 &  20.7 &  $-$11.1   & 3.78 &  2.36 &  1.99 &  1.45 &  2.91   \\    
  3C111             &  0.0485 &  13.3 &  $-$10.6 &  20.1 &  $-$10.0   & 3.33 &  1.95 &  2.89 &  1.04 &  3.81   \\    
  M87               &  0.0042 &  13.0 &  $-$10.5 &  18.5(?) &  $-$10.4   & 2.68 &  1.14 &  4.19 &  0.23 &  5.11   \\

\label{tableRadioPlanck}
\end{longtable}


\end{document}